\pdfoutput=1

\documentclass[11pt]{article}

\usepackage{geometry}
\usepackage{setspace}
\usepackage{caption}
\usepackage{graphicx}
\usepackage{csquotes}
\usepackage{booktabs}
\usepackage{tabularx}
\usepackage{listings}
\usepackage{comment}
\usepackage{enumerate}
\usepackage{float}
\usepackage{xcolor}
\usepackage{nameref}
\usepackage{hyperref}
\usepackage{todonotes}
\usepackage{makecell}

\usepackage{mathptmx}
\usepackage[T1]{fontenc}

\usepackage[style=biblatex-spbasic,
    natbib=true,
    useeditor=false,
    url=false,
    isbn=false,
    dashed=false
    backend=biber]{biblatex}
\AtEveryBibitem{%
  \clearfield{day}%
  \clearfield{month}%
  \clearfield{endday}%
  \clearfield{endmonth}%
}

\DeclareUnicodeCharacter{2121}{\textsc{Tel}}

\newenvironment{biseabstract}{%
\begin{quote} \bf}
{\end{quote}}

\newenvironment{bisekeywords}{%
\begin{quote} \it \textbf{Keywords -}}
{\end{quote}}

\title{A Systematic Review of Business Process Improvement: Achievements and Potentials in Combining Concepts from Operations Research and Business Process Management}

\author{
Michel Kunkler$^{1\ast}$, Felix Schumann$^{1}$, and Stefanie Rinderle-Ma$^{1}$\\
\\
\normalsize{$^{1}$Technical University of Munich, TUM School of Computation, Information and Technology}\\
\normalsize{Boltzmannstr. 3, 85748 Garching, Germany}\\
\\
\normalsize{\{michel.kunkler, felix.schumann, stefanie.rinderle-ma\}@tum.de}\\
\normalsize{$^\ast$To whom correspondence should be addressed}
}

\date{}

\begin{document} 

\baselineskip24pt
\baselineskip18pt


\maketitle

\begin{biseabstract}
  Business Process Management and Operations Research are two research fields that both aim to enhance value creation in organizations.
  While Business Process Management has historically emphasized on providing precise models, Operations Research has focused on constructing tractable models and their solutions.
This systematic literature review identifies and analyzes work that uses combined concepts from both disciplines.
  In particular, it analyzes how business process models have been conceptualized as mathematical models and which optimization techniques have been applied to these models.
  Results indicate a strong focus on resource allocation and scheduling problems.
  Current approaches often lack support of the stochastic nature of many problems, and do only sparsely use information from process models or from event logs, such as resource-related information or information from the data perspective.
\end{biseabstract}

\begin{bisekeywords}
Business Process Management; Operations Research; Process Improvement; Systematic Literature Review
\end{bisekeywords}



\section{Introduction}
\label{sec:introduction}
The desire to increase the efficiency of business processes is a core reason for organizations to decide to implement techniques from the Business Process Management (BPM) research field \citep{tucek_main_2015}. 
%
%
%

BPM is, by definition, concerned with taking \enquote{advantage of improvement opportunities} \citep{dumas_fundamentals_2018} and its research community has covered a broad spectrum of such opportunities.
The term \textit{process redesign} is often considered as the primary approach for \textit{process improvement}, which has led to an interchangeable use of the two terms (c.f. \citet{dumas_fundamentals_2018}). The research community has also addressed \enquote{engineering or managerial challenges} \citep{van_der_aalst_business_2013}.

Yet \enquote{BPM is by no means the only discipline that is concerned with improving the operational performance of organizations} \citep{dumas_fundamentals_2018}.
Another prominent research discipline targeting the operational performances of organizations is Operations Research (OR).
OR \enquote{is concerned with quantitative (mathematical) models and their solution} \citep{eiselt_operations_2010} and has presented methods to solve optimization problems from at least the 1940s on.
In contrast, the BPM research field is relatively new, having its roots in Workflow Management system development in the mid-1990s \citep{van_der_aalst_business_2013}.
\citeauthor{dumas_fundamentals_2018} point out that OR \enquote{is generally concerned with controlling an existing process without necessarily changing it, while BPM is often concerned with making changes to an existing process in order to improve it} \citep{dumas_fundamentals_2018}.

Nevertheless, \enquote{many process improvement problems can in fact be traced back to typical problems investigated by OR} \citep{van_der_aalst_business_2016}, and hence, some authors from the BPM research community (e.g., \cite{van_der_aalst_business_2013, van_der_aalst_business_2016, dijkman_1st_2023})) have in recent years promoted the idea of applying OR techniques on business processes.
Despite the shared goal of improving an organizations operational performances, both disciplines have typically distinct approaches to optimization as described in the following. 

\citet{hillier_introduction_2020} present a 6-step framework for solving optimization problems in OR:
In the first step, the problem is defined, and relevant data is gathered.
Secondly, a mathematical model is built, from which a solution will later be derived.
Thirdly, a solution method that calculates solutions to the mathematical model is developed.
The remaining steps are concerned with validating the solutions and integrating the solution method into the organization's operating environment.
An OR practitioner is typically concerned with the trade-off between precision, creating a model that sufficiently abstracts the problem, and tractability, assuring that a solution can be found in a feasible amount of time \citep{hillier_introduction_2020}.
When a problem appears intractable for some solution method, an OR practitioner can either change the solution method (e.g., to a heuristic) or simplify the model \citep{taha_operations_2017}.
In contrast, in the BPM research area several standardized modeling languages have been developed to depict an organization as a whole, usually divided into different perspectives, such as the control flow, resource, data, organization, or time perspective.
These modeling notations can be classified into formal and informal ones.
Formal process modeling notations have a mathematical background (e.g., can be depicted as Petri net) and can be analyzed using formal techniques. Informal notations such as BPMN do not possess the analytical capabilities but are often easier interpreted by humans. 
%
%
\citet{dumas_fundamentals_2018} distinguish between two groups of techniques that can be applied for business process optimization, namely \textit{qualitative} and \textit{quantitative process analysis} techniques:
While \textit{qualitative analysis} encompasses approaches that can be applied to gain insights into processes, these insights \enquote{are sometimes not detailed enough to provide a solid basis for decision making} \citep{dumas_fundamentals_2018}.
Conversely, \citet{dumas_fundamentals_2018} enumerate flow analysis, queuing theory, and process simulation methods as \textit{quantitative analysis} techniques.
Clearly, as OR \enquote{is concerned with quantitative (mathematical) models and their solution} \citep{eiselt_operations_2010}, the \textit{quantitative analysis} techniques in BPM are similar to OR techniques.
%


An OR project as described by \citet{hillier_introduction_2020} is a finite sequence of steps designed to serve a specific optimization goal. In contrast, a key concept in BPM is the BPM lifecycle \citep{van_der_aalst_business_2003}, which categorizes means to operationalize business processes into four recurring phases, i.e., the Process (re-)design, system configuration, process enactment, and process diagnosis phase.
The output of one phase is taken as input for the next phase and the cycle is continuously executed to reach ongoing improvement.


Existing surveys concerned with Business Process Optimization focus on the improvement of the enactment phase of the BPM life cycle. \citet{schulte_elastic_2015} have analyzed methods of resource allocation and scheduling of business process in particular for the cloud computing domain. \citet{pufahl_automatic_2021} analyze the BPM data sources that are used as input for resource allocation approaches of business processes. Here a specific focus is on the used process models and data for different allocation algorithms during enactment. 

\begin{figure}[h]
\includegraphics[width=\textwidth]{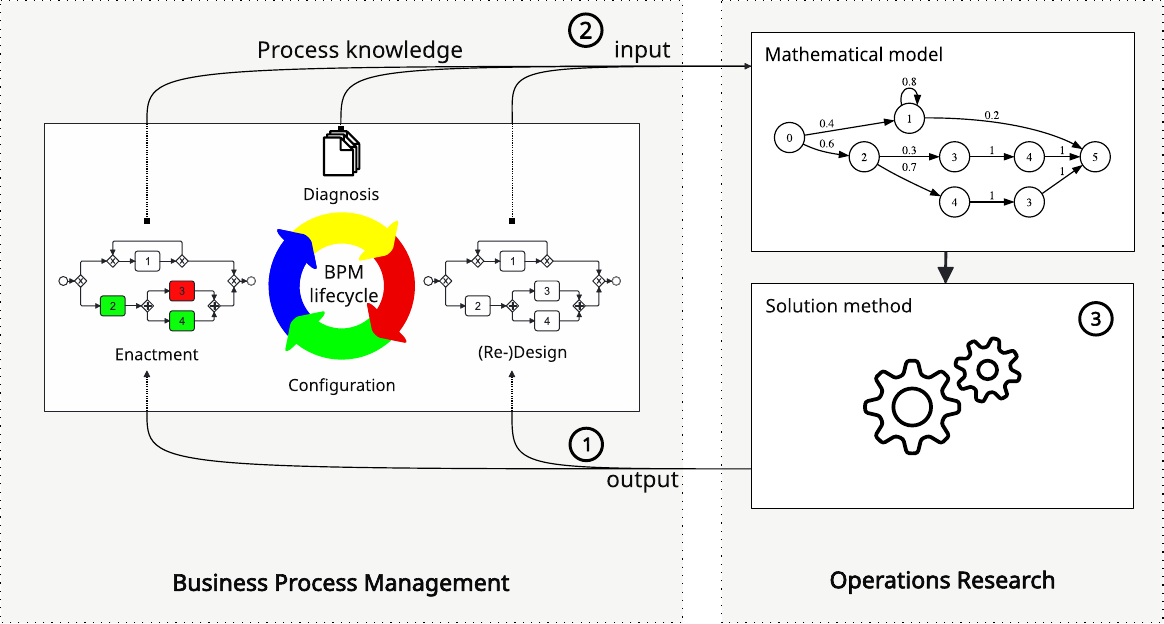}
\centering
\caption{Quantitative optimization of business processes using OR methods}
\label{fig:bpm_or}
\end{figure}

We design the integrated model shown in \autoref{fig:bpm_or}, which combines the OR-driven optimization project methodology based on \citet{hillier_introduction_2020} with the BPM life cycle. By following this model, our survey enhances existing works by considering business process optimization as a holistic, continuous effort, which can affect the (re-)design and the enactment phase. This integrated model is based on the idea in \citet{dumas_fundamentals_2018}, that \textit{quantitative analysis} methods can be applied in the (re-) design and the enactment phase \textcircled{1} based on data obtained during process diagnosis. We further enhance this definition
by also looking at data gathered during enactment and (re-)design \textcircled{2}.

This survey enhances existing works by analyzing the impact on design \underline{and} enactment phase \textcircled{1}, by analyzing all inputs derived from a process (model) used to construct the mathematical optimization model \textcircled{2} and analyzing the solution method to solve the mathematical model to optimality \textcircled{3}. 

%
%
%

The conducted systematic literature review is designed in coherence with \autoref{fig:bpm_or} and analyzes the application of OR techniques in the context of business processes improvement. 
%
Our first objective is to provide an overview of which optimization purposes, i.e. resource allocation, have been addressed with OR techniques.
Therefore, we will survey not only optimization purposes that concern the enactment phase but also the (re-)design phase (see \textcircled{1}).
We do not perceive the remaining two phases, configuration and diagnosis,  as applicable to OR techniques, as the diagnosis phase is concerned with aggregating data post-ex, and the configuration phase with implementing a process model, both bearing no potential for direct process improvements.
Our second objective is to survey how mathematical models are obtained from process data and in which form this data is structured.
We review which process related data is used for building a mathematical model. We categorize the data into data obtained from the enactment and diagnosis phase, i.e., runtime data or event logs, and data from the (re-)design phase, i.e., process models. In particular, we are interested in which process mining and modeling techniques are suitable for the creation of an OR model (see \textcircled{2}).
Our third objective is to survey the solution methods used to solve an optimization problem.
For this objective, we are interested in which solution methods appear helpful for solving mathematical models obtained from process-related data (see \textcircled{3}).

This work may contribute to the BPM community, as the results indicate what kind of process models and process data are necessary for business process optimization and discuss missing modeling standards.

This work may also contribute to the OR community, as the results indicate meaningful mathematical models and solution techniques for business process optimization.

The remainder of this work is structured as follows:
In \autoref{sec:research_method} we outline our research method for the literature review.
The findings of the review are presented in \autoref{sec:results}.
In \autoref{sec:challenges} we present identified challenges that should be considered by future research.
Related work is presented in \autoref{sec:related_work}. \autoref{sec:discussion_conclusion} discusses and concludes the systematic literature review.
\section{Research Method}
\label{sec:research_method}
A Systematic Literature Review (SLR) as a research method identifies, analyzes and interprets relevant research related to specific research questions in an unbiased way.
For conducting the SLR, we adopted the \enquote{Guidelines for performing Systematic Literature Reviews in Software Engineering, Version 2.3} by \citet{kitchenham_guidelines_2007}.
Hence, we begin by defining research questions.
Based on the defined research questions, we derive a search strategy, describe our strategy for identifying relevant works by defining in- and exclusion criteria, and present a data extraction template for extracting relevant data from the identified works.
%
%
\subsection{Research question specification}
\label{sec:research_questions}
\newcounter{RQ}
We break down the overall motivation for this study, of how OR methods can be applied in BPM, into three research questions that arise from the integrated model (\autoref{fig:bpm_or}).
The first research question addresses the optimization purposes, i.e., which optimization problems are addressed in the (re-)design and the enactment phase.
\begin{enumerate}[align=parleft]
    \label{RQ:1}
    \item[\textbf{RQ1}:] Which optimization purposes are addressed when techniques from OR are applied to business processes? 
\end{enumerate}
\textbf{RQ1} seeks to provide insights into the types of optimization problems researchers have predominantly addressed and reveal those that present promising opportunities for future research. As shown in \autoref{fig:bpm_or} a focus of this question is the life cycle phase that is affected by the optimization algorithm.

The second research question addresses the interaction between process models and their representation as formal mathematical models, which will later be used for the application of OR solution methods.
%
\begin{enumerate}[align=parleft]
    \label{RQ:2}
    \item[\textbf{RQ2}:] How are business processes modeled or mined for the application of OR techniques?
\end{enumerate}
\textbf{RQ2} aims to identify the most suitable business process modeling and mining techniques for creating mathematical optimization models. Answering this research question tries to reveal opportunities and future needs for improving business process modeling and mining techniques within the BPM research field. The analytical goal is to give insight into the currently used modeling approaches and possible standards for the different process perspectives, control-flow, resources, and data.

The third research question shifts the focus to the OR field.
It addresses the optimization techniques, i.e. algorithms, used to obtain solutions from the mathematical models:
\begin{enumerate}[align=parleft]
    \label{RQ:3}
    \item[\textbf{RQ3}:] Which solution techniques are used to optimize business processes? (3)
\end{enumerate}
Answering \textbf{RQ3} is intended to reveal mathematical solution approaches that align with business processes. The focus in answering this question should be less on naming specific algorithms but on describing the type of solutions that are delivered by the approaches, i.e., approximations or optimal solutions.

\subsection{Search Strategy}

To provide a comprehensive review, we used two literature databases, i.e., \textit{Scopus} and \textit{Web of Science}, that have been acknowledged as world-leading comprehensive databases \citep{zhu_tale_2020}.
%
%

Search queries are designed to find all relevant works for answering the posed research questions while maintaining a feasible number of results.
The detailed description of the search strings is given in \autoref{table:keywords}.
The search queries consist of two groups of keywords:
One group targets the BPM field, and the other group the OR field.
The keyword groups were joined with a logical 'AND' to target publications at the intersection of the BPM and OR fields.
\begin{table}[H]
\begin{tabular*}{\linewidth}{l}
\toprule
Search queries \\
\midrule
\midrule
Web of Science \\
\begin{lstlisting}[
    basicstyle=\scriptsize
]
TITLE-ABS-KEY (
    ("BUSINESS PROCESS MANAGEMENT" OR "WORKFLOW MANAGEMENT" OR "PROCESS MINING"
        OR "PROCESS ANALYTICS") AND
    ("OPERATIONS RESEARCH" OR "OPERATIONAL RESEARCH" OR "SCHEDULING" OR "ALLOCATION" OR
        "ROUTING" OR "OPTIMIZATION" OR "OPTIMISATION")
)
\end{lstlisting} \\
\midrule
Scopus \\
\begin{lstlisting}[
    basicstyle=\scriptsize
]
TS=(
    ("BUSINESS PROCESS MANAGEMENT" OR "WORKFLOW MANAGEMENT" OR "PROCESS MINING"
        OR "PROCESS ANALYTICS") AND
    ("OPERATIONS RESEARCH" OR "OPERATIONAL RESEARCH" OR "SCHEDULING" OR "ALLOCATION"
        OR "ROUTING" OR "OPTIMIZATION" OR "OPTIMISATION")
)
\end{lstlisting} \\
\midrule
\bottomrule
\end{tabular*}
\caption{Search strings}
\label{table:keywords}
\end{table}
%
%

The BPM keywords consist of the term \textit{Business Process Management} itself and its synonym \textit{Workflow Management}.
\textit{Workflow Management} was chosen because BPM has its roots in workflow management. Both terms have been used interchangeably (e.g. in \citet{van_der_aalst_business_2013}).
A sub-field of BPM is process mining.
It differs from traditional BPM as it does not focus on process models but on event logs from which a process model can be derived \citep{van_der_aalst_process_2012}.
The term \textit{Process Mining} is selected to incorporate works that make use of process mining techniques for OR optimization techniques. Since process mining uses data gathered during enactment it puts attention on \textbf{RQ2}.
Similarly, Business Process Analytics is a sub-field of BPM and describes \enquote{the family of methods and tools that can be applied to [...] event streams in order to support decision-making in organizations} \citep{vom_brocke_business_2010}.
%
In a test run, the term \textit{Business Process} yielded an infeasible amount of results, often unrelated to the BPM discipline, and was therefore disregarded from the search strings.

As one of the OR keywords we use term \textit{Operations Research}.
To avoid language bias, we add term \textit{Operational Research}, which is more prominent in British English.
Terms \textit{Optimization} and \textit{Optimisation} are selected to include all publications that emphasize an optimization purpose without necessarily referring to OR.
The keywords \textit{Scheduling} and \textit{Routing} describe common problem categories that have been addressed in OR publications since the 1990s (c.f. \citet{laengle_forty_2017}).
Although the keyword \textit{Allocation} has received less attention in OR in recent years (c.f. \citet{laengle_forty_2017}), it is chosen because it has also been addressed in BPM (c.f. \citet{pufahl_evaluating_2015}).


The keywords are used to search within the title, abstract, and keywords of publications.
The search was conducted in November 2023, resulting in 3252 initial studies.

\subsection{Relevant research identification}
\label{sub:relevantide}
To identify relevant works from the initial results, we apply the selection criteria shown in \autoref{table:selection_criteria}.
\begin{table}[h]
\begin{tabularx}{\textwidth}{XX}
\toprule
Inclusion criteria & Exclusion criteria \\
\midrule
\midrule
\textbullet{} peer-reviewed conference papers or journal articles in English
    & \textbullet{} books or book chapters \\
\textbullet{} publications that aim for the optimization of business processes
    & \textbullet{} publications that are concerned with grid computing, cloud computing, or scientific workflows \\
\textbullet{} publications that consider formal business process models and/or event logs
    & \textbullet{} position papers \\
\textbullet{} publications that are concerned with automatic and prescriptive optimization techniques
    & \textbullet{} manual optimization or re-engineering \\
\bottomrule
\end{tabularx}
\caption{Inclusion/exclusion criteria}
\label{table:selection_criteria}
\end{table}

%
%
Overall, we expect studies to include elements of BPM and OR in order to be deemed relevant.
Hence, we include only publications that address the improvement of business processes.
Since OR techniques necessitate a formal mathematical model, we include only publications that either explicitly consider process models with a formal background (e.g., Petri nets) or studies that create mathematical models based on data from event logs.
Because OR is concerned with quantitative solution techniques, we include only works that present solution approaches that lead to an automatic improvement with prescriptive outcomes.
Conversely, studies that present non-quantitative solution approaches, i.e., guidelines for manual process optimization, and non-automatic improvement approaches, i.e. applied manual process design from interviews, are excluded.


%
%
After a test run, we noted that the keyword \textit{workflow management} is heavily used in the field of scientific and cloud computing.
In these fields, workflows are often referred to as \textit{scientific workflow} or \textit{grid workflow}.
Studies in these fields are concerned with different problem statements than publications in the BPM domain.
Workflows from the scientific and cloud computing domain deal with problems centered around \enquote{the transportation and analysis of large quantities of data} \citep{barker_scientific_2008}, while \enquote{business workflow tools look more like traditional programming language} \citep{barker_scientific_2008}.
While imperative business process models typically make use of complex workflow patterns (c.f. \citet{van_der_aalst_business_2003}), the scientific and cloud computing domain has spent less attention on complex workflow patterns, i.e., scientific processes are often only depicted as directed graphs and therefore not relevant to this study.
Due to the diverging challenges of both disciplines, we decided to exclude works that address the cloud, scientific, or grid computing domains.

%
The inclusion and exclusion criteria are applied to titles, keywords, and abstracts of the publications that our search yielded and subsequently rated by the authors on a Likert-5 scale \citep{likert_technique_1932}. All publications that are ranked better than 3 are chosen for reading the full text.
\\
The initial search yields $3252$ publications. After removing duplicates, $2304$ publications remain for the title, keywords, and abstract screening. 
After the categorization along the inclusion and exclusion criteria, $176$ publications remain with a 4 or 5 rating on the Likert scale and are consequently selected for full-text analysis.
Full-text copies are obtained for $170$ publications as six publications could not be found online.
After reading the full texts, another $88$ studies are excluded due to not meeting the inclusion criteria.
A final number of $82$ publications are considered as relevant for further assessment.
%
%
%

%
%
%
%
%
%
\subsection{Data extraction}
\label{sec:data_extraction}
Following the full-text review of the $82$ publications identified as relevant in \autoref{sub:relevantide}, we extract data using the data extraction form shown in \autoref{table:data_extraction}.

\begin{table}[H]
\begin{tabularx}{\textwidth}{lll}
\toprule
Data extraction & Evaluation Criteria & Categorisation \\
\midrule
\midrule
RQ 1: Optimization purpose & Effected BPM life-cycle phase & [Enactment, Design] \\
& Problem category & [8 Categories, see \autoref{sec:data_extraction}    ] \\ 

RQ 2: Optimization input  & Process model & [Yes, No] \\
& Workflow pattern & \makecell[tl]{[Sequential, AND, XOR, \\ ~Deferred Choice]} \\
& Event log & [Yes, No] \\
& Multiple process instances & [Yes, No] \\
& Resource modeling & [Yes, No] \\
& Resource Representation & [pooled, individual, both] \\
& Resource Performance Indicators (RPI) & [None, 1 RPI, complex] \\
& Resource defines task duration & [Yes, No] \\
& Time profiles & [Yes, No] \\
& Data perspective & [Yes, No] \\
RQ 3: Solution method & Solution approaches & \\
& Optimality & [Yes, No]  \\

\bottomrule
\end{tabularx}
\caption{Data extraction}
\label{table:data_extraction}
\end{table}

The data extraction form is supposed to ensure a systematic approach to answering the research questions.
To answer \textbf{RQ1}, publications are categorized on whether their proposed process optimization approaches affect the design phase, e.g., planning an optimal amount of resources prior to a process's enactment, or the enactment phase, e.g., allocating resources to process activities during its enactment.
To provide a detailed analysis, we define eight optimization purpose categories.
The initial categories are discussed during the review protocol design and refined after all relevant publications were read. The categories from the OR side are derived from typical problems that are addressed in OR projects \citep{hillier_introduction_2020}, i.e., scheduling, batching, and resource planning. Problems typically addressed by BPM are concerned with changing the process model \citep{dumas_fundamentals_2018}. Thus, we decide on the following categories, i.e., selection of the best process model, control-flow re-engineering, process navigation, and action recommendation. The problem of resource allocation is addressed by both, i.e., BPM and OR research. We add the \textit{SLA planning} category as one publication could not be sorted into other existing categories. Consistently, categories that have been defined upfront, such as routing, are dropped due to no fitting publications.

To answer \textbf{RQ2}, concerned with how mathematical optimization models are obtained, we create ten extraction fields.
First we review whether a process model and/or an event log is used to obtain an optimization model.
If a process model is used as input for the optimization model, we review which workflow patterns are supported in the optimization model.
%
When a publication models the resource perspective, we extract how the resources are represented, i.e., as individual resources and/or as pooled resource groups, i.e., organization groups.
If resource modeling is done, the representation of resource performance indicators (RPI) is checked. Such performance indicators comprise execution times and workload.
If performance indicators are modeled, it is distinguished if only one measure is covered (1-RPI) or if multiple performance indicators are considered.
It is also analyzed if a resource is enabled to actively influence the execution time of tasks. This describes the connection between control-flow and resource perspective.
To better categorize scheduling and resource allocation approaches, the presence of time profiles (e.g. availability calendar) is marked.
Additionally, it is recorded whether the data perspective is employed in the optimization model construction.
Finally, we identify if the optimization model considers multiple, possibly concurrently running, process instances.

To answer \textbf{RQ3}, concerned with the optimization techniques used for solving the mathematical models, we focus on the used procedures.
Since publications often combine several algorithms to obtain a solution, we do not use a structured field to extract the proposed solution approaches.
We extract whether the applied method finds an optimal solution to the mathematical model or an approximation and how this solution is found, e.g.,  by heuristics.
\section{Results}
\label{sec:results}
This section provides an overview of the selected publications and presents the analysis results with respect to research questions \textbf{RQ1}--\textbf{RQ4}. The full research protocol is provided as supplementary material\footnote{https://anonymous.4open.science/r/SLR\_OR\_BPM-B261/}.
%
%
%
%
%
%
%
%
%
%
%
%
%
\subsection{RQ1: Addressed optimization problems}
\label{sec:problem_categories}
To answer \textbf{RQ1}, we first categorize the optimization purposes of the selected publications into 
design or enactment phase.
Publications categorized into the design phase are concerned with decision-making problems that occur prior to enacting a business process and affect the planning of its properties, for example, planning the control flow perspective or planning the required amount of resources \underline{before} starting process enactment.
Conversely, publications categorized into the enactment phase are concerned with decision-making problems that occur during the execution of a business process such as allocating resources to a business process task for its execution.
%

\begin{figure}[h]
    \centering
    \includegraphics[width=\textwidth]{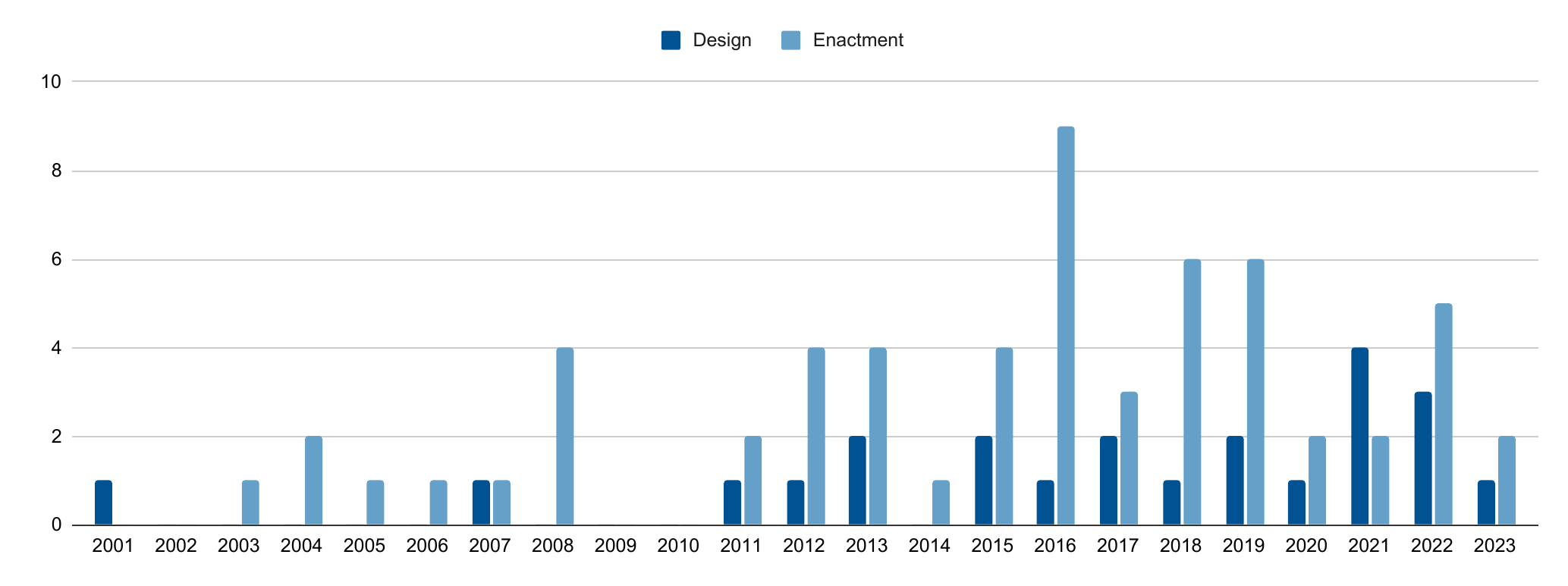}
    \caption{Publishing year and life cycle phase}
    \label{fig:timeline}
\end{figure}

The majority of the publications (60, 72\%) addresses optimization problems in the enactment phase, while 23 (28\%) publications address the design phase. \autoref{fig:timeline} shows the distribution of the selected studies on a timeline. 
The majority of publications addressing quantitative optimization methods in BPM have been published since 2011.
The number of publications that address optimization in the enactment phase has peaked in 2016 and the number of publications that address optimizations in the design phase has grown since 2021. An interpretation might be that existing enactment optimization problems are well-understood while design optimization problems are currently gaining momentum with more powerful algorithms to include model aspects into the optimization.

%

\begin{figure}[h]
    \centering
    \includegraphics[width=0.75\textwidth]{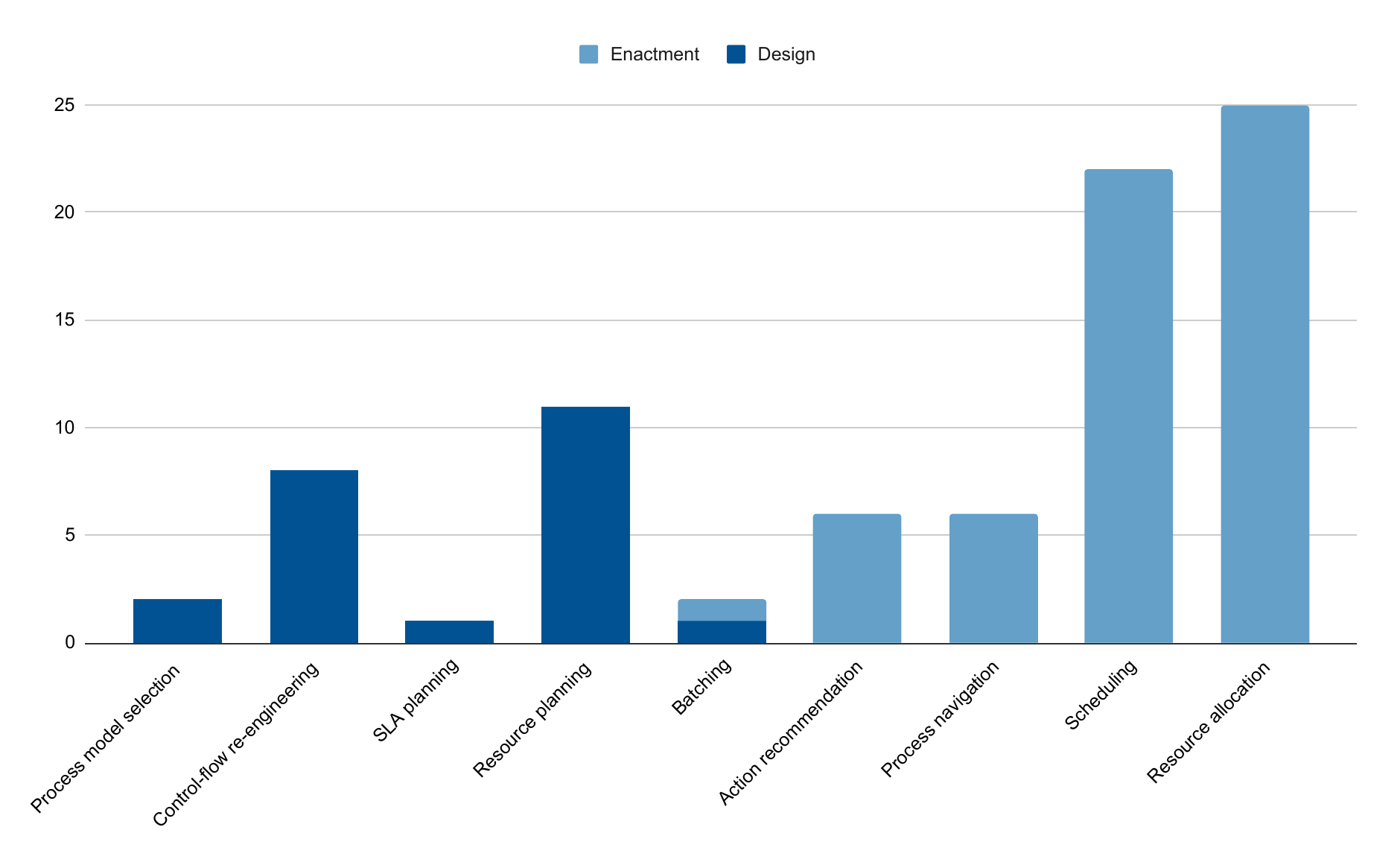}
    \caption{Problem categories and life cycle phases}
    \label{fig:problem_tree}
\end{figure}

The addressed optimization problems can be subsumed by nine problem categories. As described in \autoref{sec:data_extraction}, initial problem categories were discussed among the authors during the creation of the review protocol.
The final set of nine problem categories was established after having finished reading all selected publications. Four categories are addressed in the design phase, i.e., \textit{process model selection}, \textit{control flow re-engineering}, \textit{service-level agreement (SLA) planning}, and \textit{resource planning}, and 
four categories are addressed in the enactment phase, i.e., 
\textit{Action recommendation}, \textit{process navigation}, \textit{scheduling}, and \textit{resource allocation}.
Category \textit{batching} is addressed in both phases.
\autoref{fig:problem_tree} shows the number of publications for each of the problem categories.
The majority of publications is concerned with \textit{scheduling} and \textit{resource allocation} in the enactment phase, and with \textit{control flow re-engineering} and \textit{resource planning} for design time. In the following a description of the individual problem categories and their publications is presented.
%
%
%
%

%
%


%
%
%
\textbf{Process model selection} comprises the two works of \cite{satyal_ab_2018, satyal_ab-bpm_2017}, which are concerned with finding the best-performing process model from a set of feasible process models.
%

%
\textbf{Control flow re-engineering} publications present quantitative optimization approaches for re-engineering the control flow of a process.
\citet{wang_trust-based_2013, niedermann_design-time_2010} propose techniques to parallelize activities in an automated way to speed up the process execution.
Conversely, \citet{xu_incorporating_2013} propose an automatic approach to sequentialize activities, such that an allocation algorithm is prevented from assigning an unsuitable or expensive resource to an activity when a well-suited resource is unavailable because it is executing a parallel activity.
Two publications are concerned with (re-)engineering decision points:
\citet{bolsinger_process_2015} aims to find optimal conditional values for exclusive choices, while \citet{subramaniam_improving_2007} aims to move the decision points to the earliest possible points to enhance planning certainties.
\citet{van_der_aalst_re-engineering_2001} investigates strategies to find an optimal task order in knock-out processes where the outcome is, e.g., to approve or disapprove a loan.
\citet{lee_clustering_2017} suggest splitting a process model into a subset of process models so that the process models have a minimal inter-process coupling.
These smaller processes can be used individually for further scheduling or allocation techniques.
Lastly, \citet{song_identifying_2022} presents an approach that finds the minimal number of change operations needed to transform a control flow model into another one.

%
\textbf{SLA planning} comprises the work of \citet{cho_new_2017} concerned with an automatic approach for proposing SLAs based on business processes.
%

%
\textbf{Resource planning} publications aim to plan either the number of resources needed for process execution or the properties of resources such as their locations.
Finding the optimal number of resources is the goal of \citet{peters_resource_2021, prokofyeva_clinical_2020, antunes_solution_2019, li_mining_2021, liu_workflow_2012}, while finding optimal properties is the objective for \citet{kinast_combing_2022, halawa_integrated_2021, rismanchian_process_2017}.
\citet{choueiri_discovery_2021, lopez-pintado_business_2022} propose approaches that can be used to evaluate the effect of replacing a resource with another one.
\citet{kumbhar_digital_2023} is concerned with identifying bottleneck resources.
%

%
\textbf{Batching} addresses the problem of finding optimal batch sizes for activities.
It is the only category considered in the enactment and design phases.
\citet{pufahl_evaluating_2015} focus on a static batch size, determined during design time.
In contrast, \citet{pflug_application_2016} introduce a dynamic method that adjusts the batch size during the enactment phase.

%
%
%
%
\textbf{Action recommendation} refers to the execution of a next task in a process instance or for a resource or to initiating a new process instance.
\citet{barba_user_2013} recommend the most suitable next activity to a resource, i.e., a user.
The pool of possible activities can stem from different process instances with different priorities or service-level agreements that must be obeyed.
\citet{weinzierl_prescriptive_2020} recommend to skip process activities, e.g., when a deadline can only be met if checking the credibility of a customer is skipped.
\citet{banaszak_workflows_2003} recommend whether a new process instance should be spawned or not, i.e., an offer be accepted or not.
\citet{kurscheidt_netto_enabling_2021, ruschel_establishment_2020, ruschel_mining_2017} focus on recommending time points for maintenance tasks.
%

%
\textbf{Process navigation} proposes a viable process path, e.g., a set of activities required to complete a process instance.
\citet{medeiros_constraint_2017, julia_real_2008, julia_p-time_2004} are primarily concerned with (resource) deadlocks that might occur during the process executions and suggest identifying feasible execution strategies that are not exposed to deadlocks by using Monte Carlo simulation.
\citet{comuzzi_ant-colony_2019, zheng_enterprise_2018} 
present approaches to select the optimal branch at deferred choices
Similarly, \citet{bando_automatic_2022} conceptualize robot movements as a process model and address the optimal navigation through the process model.
%

%
%
\textbf{Scheduling} comprises approaches that schedule the execution of business processes in advance and in doing so, typically also plan the allocation of resources to tasks.
\citet{yaghoubi_tuning_2018, xu_resource_2016} focus on instance-level scheduling, i.e. planning the starting times of process instances.
While \citet{yaghoubi_tuning_2018} aims to balance the resource workload, \citet{xu_resource_2016} aims to maximize the number of instances that can be finished in a given timeframe.
%
The other publications focus on scheduling at task level, i.e., they consider a process model for scheduling the execution of activities and propose either to solve constraint programming (CP) formulations \citep{senderovich_learning_2019, jimenez-ramirez_generating_2013, barba_automatic_2013, avanes_adaptive_2008}, Answer Set Programming (ASP) formulations \citep{havur_benchmarking_2022, havur_history-aware_2019, havur_resource_2016}, or (mixed) integer linear programming ((M)ILP) formulations \citep{guastalla_workshift_2023, di_cunzolo_combining_2023, low_revising_2016, bae_planning_2014, hirsch_optimization_2012}.
\citet{reveliotis_real-time_2016} formulate the scheduling problem as markov reward process and propose to find an approximate solution via stochastic approximation methods.
\citet{ng_business_2018,low_revising_2016} propose to use meta-heuristics,
\citet{hsieh_hybrid_2017, hsieh_location-aware_2016} multi-agent systems, and \citet{senderovich_conformance_2016, cho_evidence-based_2019} propose to incrementally build schedules via heuristics.
%

%
\textbf{Resource allocation: } 
In \citet{djedovic_optimization_2016, djedovic_innovative_2018, bai_risk_2013}, resources are allocated to one specific type of activity.
The other publications are concerned with allocating resources to tasks, i.e., instantiated activities:
In \citet{liu_multi-level_2022,yeon_experimental_2022,lee_dynamic_2019, arias_human_2018,ismaili-alaoui_business_2018, bae_--fly_2015,liu_q-learning_2015,yang_approach_2012,liu_semi-automatic_2008,zhao_entropy-based_2016,zhao_optimization_2015, alonso_workflow_2007}, allocation decisions are based on the best matching of properties of tasks to resources or resources to previous resources.
Similarly, \citet{xu_performance_2012, combi_task_2006, zeng_effective_2005, lin_practical_2004} propose allocation heuristics.
\citet{delias_formulating_2023,delias_joint_2008} propose approaches to minimize resource conflicts and maximize the active resources.
\citet{pla_multi-attribute_2012} propose an allocation approach based on multi-agent auction.
Using reinforcement learning for training allocation policies is proposed in \cite{neubauer_resource_2022, huang_reinforcement_2011}.
Finding the best matching among the available resources and unassigned and future tasks by formulating the problem as minimum-cost maximum-flow problem is presented in \cite{park_prediction-based_2019}.
\citet{arias_human_2018} propose two resource criteria driven allocation approaches, i.e., immediately allocating a task to the resource with the best resource criteria matching, and allocating a set of tasks (batch) to resources such that the best matching for all tasks is achieved.


%
%
%
%
\subsection{RQ 2: Business process modeling and mining}
%
%
%
%
%
Finding an appropriate mathematical model that precisely depicts reality while maintaining computational tractability is key to solving an OR problem. Hence, in the following, we will investigate how business processes are modeled or mined in order to obtain mathematical optimization models that are later solved through various solution techniques. This section starts with an overview of the data sources.
Afterwards, we investigate how the individual process perspectives, i.e., the control flow, resource, and data perspective, are modeled and how their data is used to obtain optimization models.
%
%
%
%
\subsubsection{Data sources}
Information from the control flow perspective, represented by process models, is used in 49 (59\%) publications.
A process model is either assumed to be known or is first mined via process discovery techniques.
Process mining for data gathering is explicitly used in 43 (41\%) publications.
%
Information about resources is used in 70 (84\%) publications.
Nine (11\%) works consider data from the data perspective. 
Some publications do neither mathematically model the control-flow perspective, nor the resource perspective, and, e.g., only use runtime data to obtain solutions (e.g., \cite{liu_semi-automatic_2008}).
%
%
%
\subsubsection{Control flow modeling}
%
%
Overall, $49$ publications mathematically formalize the control flow perspective; $46$ consider imperative process models, and three consider declarative process models.
%
%
In order to assess the expressive power of the proposed models and associated mathematical formulations, we base our analysis on selected control flow workflow patterns \citep{van_der_aalst_workflow_2003}, i.e., sequence, parallel split and synchronization, as well as exclusive choice and merge.

%
%
Workflow patterns are conceptualized in two distinct ways:
Either they are formalized as constraints into mathematical models from which solutions are obtained directly by using \textit{analytical optimization} techniques. 
The sequential pattern, for example, can be formalized as a constraint in a constraint programming formulation such that no activity may be scheduled before its predecessor has been finished.
The second approach is to conceptualize workflow patterns as constraints in simulation models, e.g., the simulation model only allows an activity to be scheduled after its preceding activity has been finished. A solution, typically a policy, might then generate solutions that do violate the constraints, leaving the simulator to decide how to handle these situations.
Good policies can then be trained using \textit{simulation optimization} techniques \citep{amaran_simulation_2016}.
Of the 49 publications that formalize the control-flow perspective for optimization, 19 (39\%) use \textit{simulation optimization} techniques and 30 (61\%) publications use \textit{analytical optimization} techniques.

%
%
All of the 46 publications that are concerned with \textit{imperative process models} incorporate the sequential pattern in their analytical or simulation models.
38 publications additionally conceptualize and consider information from parallel splits and their synchronizations.
However, only 17 publications consider exclusive choices and their merges.
Interestingly, 12 publications consider the deferred choice pattern, a workflow pattern at which the environment, e.g., a business process management system, can choose a path to be taken \citep{van_der_aalst_workflow_2003}.
For publications that address optimizations in the enactment phase, the deferred choice pattern is conceptualized even more often than the exclusive choice pattern, as shown in \autoref{tab:wf_patterns}.
%

\begin{table}[!ht]
    \centering
    \begin{tabular}{|l|c|c|}
    \hline
        ~ & Affects Design Phase & Affects Enactment Phase \\ \hline
        Sequential & 13 & 33 \\ \hline
        AND & 10 & 28 \\ \hline
        XOR & 8 & 9 \\ \hline
        Deferred Choice & 1 & 11 \\ \hline
    \end{tabular}
    \caption{Workflow patterns considered}
    \label{tab:wf_patterns}
\end{table}

\subsubsection{Exclusive choices}
%
%
Exclusive choices are points \enquote{in the workflow process where, based on a decision or workflow control data, one of several branches is chosen} \citep{van_der_aalst_workflow_2003}.
Exclusive choices are interesting because, unlike AND-splits, where all subsequent branches must be executed,
the branch chosen at an exclusive choice can typically not be foreseen during design time or at model instantiation.
%

%
%
Exclusive choices are themselves subjected to optimization in two publications.
In \citet{subramaniam_improving_2007}, data-flow analysis is conducted to move decision points to the earliest possible points to reduce the number of possible paths of a process as early as possible.
In \citet{bolsinger_process_2015}, it is assumed that the conditions of exclusive choice are not ideal and, therefore, subjected to optimization to maximize the value contribution of a process.

%
%
In the context of scheduling and planning, 10 publications conceptualize exclusive choices for the application of \textit{analytical optimization} techniques, while 5 for \textit{simulation optimization} techniques.
Analytical approaches most commonly use probabilistic techniques, including \citet{havur_benchmarking_2022, peters_resource_2021, havur_history-aware_2019, bolsinger_process_2015, delias_optimizing_2011, bai_risk_2013} which use transition probabilities for the individual branches from an exclusive choice gateway.
%
\citet{barba_automatic_2013} schedule the execution of a business process until an exclusive choice is reached.
\cite{havur_history-aware_2019} combine the idea of rescheduling and using transition probabilities. They propose considering the most likely branch of an exclusive choice when it exceeds a predefined threshold.

In resource planning for emergency processes, the work of \citet{li_mining_2021} considers the worst-case branch for estimating the maximal number of simultaneously required resources. Another non-probabilistic approach is presented in \citet{choueiri_discovery_2021}.
Under the assumption that resources affect the data perspective of processes, their approach finds associations between resources and the control flow path.

%
Exclusive choices in \textit{simulation models} are considered in the works of \citet{yaghoubi_tuning_2018, lin_practical_2004, lopez-pintado_business_2022, choueiri_discovery_2021, djedovic_innovative_2018, liu_workflow_2012, guastalla_workshift_2023}:
Only \citet{djedovic_innovative_2018} describe how they set up their simulation model, e.g., how they chose probability distributions for modeling task durations.
The other publications lack detailed descriptions about how they set up their simulation models.
%
\subsubsection{Resource modeling}
\label{sec:resource_modeling}
%

A key approach to business process improvement is to optimize the planning and allocation of resources that execute the tasks in a process. Following the argumentation by \citet{peters_resource_2021}, the performance of a process can be improved by lowering the waiting times until a resource is available (tactical allocation), which can be achieved by having more resources available. The performance of a process can also be improved by allocating its tasks to more effective resources and thus improving task execution times (operational allocation). Looking at the BPM life cycle, tactical resource allocation is an optimization affecting the design phase (e.g., making more resources available), and operational resource allocation affects the enactment of a process (e.g., allocating resources based on performance). Our study identified 70 papers that actively model resources and use them as input to solve their mathematical model. 

Different information on resources and their performance is needed to make better decisions when allocating tasks to certain resources. We defined the categories shown in \autoref{table:data_extraction} to analyze how resources are modeled throughout different approaches. Besides the overall modeling of resources, we also consider how resource performance indicators (RPI) are modeled as described in \autoref{table:data_extraction}. We check if none, one, or multiple RPIs are modeled for a resource. A simple RPI measures, for example, the execution time of a certain resource. More complex RPIs could additionally model a resource's workload, uptime, costs, and other performance indicators. We further investigate if resources are considered as individual instances or groups (i.e., organizational units), if their availability schedules (time profiles) are modeled (i.e., whether they follow a shift plan), and if their properties have an impact on task execution performance (i.e., resource \texttt{A} executes a certain task quicker than resource \texttt{B}). The following provides an overview of the different modeling choices along the (re-)design and enactment phase categorization. Due to the number of publications (70), we use selected publications as examples. The full list of publications that fit into which category can be found in the appendix.

\noindent\textbf{Design phase} 16 of the 70 publications that model resources affect the design phase with the optimization. A resource is modeled as an individual instance in 6 of these publications. In \citet{lopez-pintado_business_2022}, the difference between modeling resources as individuals and modeling resources as groups is assessed for simulation models and finds that the individual modeling of each resource leads to simulation models closer to actual event logs. In terms of modeling performance indicators, 5 publications model multiple RPIs, and 3 publications model a single performance measure. 8 publications fully disregard the performance of resources. We can see that for optimization at design time, the focus is not on resource performance, and a link to the later enactment is not always given. Yet, 7 publications consider the impact of a resource on the execution performance of a task. A majority of the publications where optimization affects the design phase model resources in a pooled manner (10). This means that multiple instances of resources are grouped, e.g., by their role, and their properties are aggregated over their group. Individual modeling is used in 5 studies. 
In terms of time profiles, only \citet{antunes_solution_2019} model a notion of a schedule in the design phase. Here the waiting times in an emergency department are optimized by improving the overall schedule design for the emergency department.

\noindent\textbf{Enactment phase} 55 publications consider resources as input for an optimization model to improve the enactment of business processes. Here, most approaches model multiple resource performance indicators (21) or at least one RPI (20). The performance is fully disregarded in 14 publications. To model these performance indicators, different modeling approaches are chosen. An advanced one is described in \citet{arias_human_2018}. The authors employ the structure of resource process cubes, which are inspired by typical online analytical processing (OLAP) technology, where each dimension represents one RPI of a resource. Using this approach, aggregations of resource-related data are easily achieved. In modeling resources itself, most approaches (41) use individually modeled resources. In 9 publications, resources are pooled, and 5 publications offer a hybrid approach that combines individual and pooled modeling. \citet{havur_resource_2016,lin_practical_2004} offer both modelings and enable an aggregation of performance measures for a group, if needed. For enactment, a total of 11 publications model time profiles as part of their resource modeling. Many of those publications are concerned with scheduling especially in the healthcare domain \citep{di_cunzolo_combining_2023,guastalla_workshift_2023,cho_evidence-based_2019}

Besides modeling resource performance, an interesting question is whether the impact of a resource on task execution performance is considered. Here, 35 publications disregard an influence completely, and 35 publications consider the influence of a resource on task execution performance. Interestingly, some works that model resources with complex properties or one performance property still disregard the impact of such properties at task level. Another view on the impact of a resource on process execution is given in \citet{huang_reinforcement_2011, choueiri_discovery_2021}. Here, the effect of a resource on the taken process path is studied.

\subsubsection{Data modeling}
\label{sec:data_modeling}

Data flow is seldomly used as input for the mathematical optimization model. An exception is \citet{subramaniam_improving_2007} where data dependency graphs are modeled. In \citet{bolsinger_process_2015}, information from the data perspective is used to improve the decision at XOR-Gateways. \citet{havur_history-aware_2019} use the data perspective to decide if a fragmentation for better allocation should be done at a decision point or not. 
Yet, the data flow information is usually highly aggregated up to a binary level (data available or not). For task prediction with XOR-Gateways, the data perspective might become more important in the future. 
%
%
%
%
%
%
%
%
%
%

%
%
%
%
%
\subsection{RQ3: Solution techniques}

%
While some publications emphasize setting up a simple formal model on which the application of optimal solution techniques is tractable,
other publications have presented complex mathematical models to which only approximate solution methods can be applied.
\subsubsection{Optimal solutions}
%
Optimal solutions are preferred when the search space is small, mostly due to a small number of input variables from the model or due to a small number of objective variables with a small value range.
The first is the case when information is strongly aggregated.
\citet{bai_risk_2013}, for example, use risk and cost information of each task as input for allocating control resources that reduce overall costs, to tasks.
The latter is the case when, e.g., only starting times of instances are to be scheduled, instead of scheduling all tasks from all instances \citep{yaghoubi_tuning_2018}.
Some publications propose optimal solutions, even when the algorithmic complexity is NP-hard.
\citet{havur_benchmarking_2022, havur_resource_2016} propose to use logic programming and \citet{senderovich_learning_2019} constraint programming to find optimal schedules for imperative process models.
Similarly, \citet{barba_automatic_2013,jimenez-ramirez_generating_2013} suggest using constraint programming for scheduling on descriptive process models.
\subsubsection{Approximate solutions}
%
Approximate solutions are the preferred solution method when the search space is intractably large or when the problem is subject to uncertainties under which an optimal solution seems to be infeasible.
Approximate solutions are especially present in publications that relate to resource allocation and scheduling, resource planning, or process navigation.
Linear programming relaxation is used for resource allocation in \citet{delias_joint_2008, delias_optimizing_2011}.
\citet{bae_--fly_2015} argue against scheduling allocations in business processes due to the problem complexity and that \enquote{various unpredicted factors} can occur during the execution. Therefore they propose to conduct on-the-fly resource allocation by matching resource properties to task properties.
Other approaches propose the use of meta heuristics.
\citet{comuzzi_ant-colony_2019, ng_business_2018} propose ant-colony optimization for process navigation and scheduling, respectively.
\citet{djedovic_optimization_2016} propose a genetic algorithm for planning the number of resources for each activity.
\citet{bolsinger_process_2015} propose an evolutionary algorithm to find good conditional values for decision points.
Overall, meta heuristic approaches find their solution by using simulation-optimization techniques that require a simulation of the process.
However, most approaches do not describe in detail how complex the simulation models become.
%
%
%

\section{Open challenges and research directions}
\label{sec:challenges}

During the SLR, we identified several research gaps, which have still not been extensively studied when combining BPM and OR. These research gaps focus on different parts of an optimization project and range from the optimization purposes itself, over the modeling of business process-related data to the solution techniques for the mathematical model. In the following, we discuss each research gap together with publications that represent a starting point to advance the research in that direction.
\subsection{Optimization purposes}

\noindent\textbf{Challenge:} Existing optimization approaches affect either the design or the enactment phase of the BPM life cycle.
Some resource planning approaches test if adding a resource enhances the process performance by employing a simulation model that uses an allocation algorithm \citep{antunes_solution_2019}. However, during enactment, neither the quality of the found plan is determined nor the simulation model is updated. 
%
Future holistic approaches should combine optimization purposes from the design, as well as the enactment phase, enabling optimal resource allocation through better resource planning and enabling better resource planning through interaction with the previous resource allocation outcome. 

\noindent\textbf{Research Direction:} An approach to create an optimization cycle that affects both, the enactment and design phase has not been provided yet and a holistic approach that applies multiple optimization purposes successfully along the BPM life cycle is missing. The approach by \citep{antunes_solution_2019} can serve as inspiration on how to evaluate the quality of a plan. To achieve optimization along the whole lifecycle, a feedback strategy between the design and enactment phases should be researched.

%
%
%
\subsection{Process models}
\noindent\textbf{Challenge:} Information provided by the process model currently finds only limited use for decision making.
Especially in the context of resource allocation and scheduling, 33 of 59 publications consider information from process models, and only 10 consider basic workflow patterns, exclusive choices, and merges.
From the 10 publications that consider exclusive choices, only \citet{havur_benchmarking_2022, havur_history-aware_2019} propose to build schedules beyond the points of exclusive choices.
However, these schedules are only built when a branch is expected to be selected with a likelihood that surpasses a predefined threshold.
No work conceptualizes exclusive choices such that stochastic optimization techniques can be applied for an analytical solution approach, e.g., for stochastic linear programming.
%
%

\noindent\textbf{Research Direction:} Optimization approaches often disregard exclusive choices in process models and the stochasticity of the problem is disregarded. \citep{havur_resource_2016,havur_benchmarking_2022} present possibilities to solve this issue, that are not generally applicable yet. Simplifying the transfer from process models to optimization models will embed optimization during process execution
\\
\noindent\textbf{Challenge:} Conceptualizing the control-flow perspective as a simulation model is another approach pursued in some publications that can be used, e.g., for what-if analysis or to find solutions to combinatorial optimization problems via simulation optimization techniques.
Most surveyed publications that use simulation models do not describe in detail how a simulation model is built from a process model or an event log.
Creating simulation models is generally considered complex and time-consuming (cf. \citep{juan_review_2015}).
If a simulation model does not depict reality, a solution based on the simulation model will never be applicable in reality.
Some approaches circumvent the issue by, e.g., suggesting to replay event logs instead of building complex simulation models \citep{van_der_aalst_change_2015}.

\noindent\textbf{Research Direction:} The automatic creation of simulation models and how to ensure a good quality of such a simulation model is an open question. Recent approaches combining log-replay methods (e.g., \citep{camargo_learning_2019}) with simulation methods to build hybrid simulation models appear promising to ease the simulation model-building process.

%
%
\subsection{Resource modeling}

\noindent\textbf{Challenge:} When modeling resources, a key question is the representation of process-related resource performance data, i.e., task performance when being executed by a given resource. The representation of such performance measures is described in \autoref{sec:resource_modeling}. In optimizing approaches for both, the design and the enactment phase of business processes, the resource perspective is still often disregarded. This could be rooted in the fact that no consensus is found on how to model resource-related data for a process. 
The publications studied in this work mostly rely on data tailored to the specific optimization use case, and thus, a variety of performance data representations is used. One of the few structures employed by multiple works is the resource profile. One implementation of resource profiles is described by \citet{arias_human_2018} in which resource profiles are represented as part of a resource process cube, which allows for OLAP operations to analyze the data. This enables multiple performance measures for each resource, unit-level characterizations, and a mapping for process-specific measures. This data structure gathers the performance of each resource for a specific task in a specific process and, therefore, enables the variable usage of any data object to improve the solution of the optimization algorithm. 


In addition, parts of the resource data are disregarded for the optimization model by many approaches, for example, the relationship between the executing resource and the performance of a task, i.e., resource \texttt{A} executes a certain task  quicker than resource \texttt{B}. Overall, current process optimization approaches lack a sophisticated analysis of the interaction between resource and task performance and, even further, resource and process performance. If BPM wants to become more proactive, as advocated in \citet{poll_process_2018}, such information is needed for reliable forecasting and better optimization models,  
and enables allocation algorithms to outperform static approaches as shown in \citet{bae_--fly_2015}.

Shifting the focus from optimization to simulation models, \citet{van_der_aalst_business_2015} further supports this research need by presenting the missing link from resources and resource data for simulation models. A generally agreed-on resource representation will help to create algorithms that incorporate multiple kinds of information into optimization and simulation approaches and enable general solutions, which can be easily adapted and reused by multiple researchers. 

\noindent\textbf{Research Direction:} A standardized modeling approach for resources, which connects resource data and process execution data, does not exist. This hampers the creation of general optimization and simulation models. \citet{arias_human_2018} provides process cubes as a possible useful representation of resource profiles. Analyzing resource performance based on their performance during processes opens the possibility of process optimization by optimizing the usage of resources instead of only optimizing the control flow.  


\noindent\textbf{Challenge:} The necessity for a standardized resource data structure accounts mainly for cases in which the required data is available or is actively recorded for all resources. Other research, such as \citet{lin_practical_2004,havur_resource_2016}, focuses on how to deal with missing resource data and uses the differentiation into pooled (grouped) resources and individual resources. 
While \citet{lopez-pintado_business_2022} show, that an individual representation can lead to improved resource allocation, a pooling of resources can be used to account for missing data. The six approaches that propose and evaluate a hybrid format between pooled resource representation and individual resource representation pioneer a promising path for future work. In \citet{lin_practical_2004, havur_resource_2016}, resources can be modeled both individually or as part of a resource pool. In terms of performance measures, if no individual record is available for a resource, the aggregated record of resources with the same role in a pool can be used to describe the expected performance of an individual resource. Further, \citet{lin_practical_2004} argue that the introduced scheduling model can better adapt to parallel machine scheduling problems by creating this hybrid modeling approach. 

\noindent\textbf{Research Direction:} Dealing with missing data for individual resources is a problem when using resource data for optimization, not only in BPM applications. \citet{lin_practical_2004} propose hybrid approaches combining individual and pooled resource data to create this missing data. In turn, more data for advanced optimization techniques is available.

\subsection{Solution techniques}
\noindent\textbf{Challenge:} Current approaches often do not model stochastic properties sufficiently, due to two reasons.
Firstly, the set of tasks required for finishing a process instance is often assumed to be known upfront.
Model-induced uncertainty, e.g., due to exclusive choices, is often not considered.
Secondly, task durations are often assumed to be deterministic, and solution methods work only with their expected values.
As real-life problems often have stochastic properties and uncertainties, it appears worth exploring the potential for applying stochastic optimization methods.
Currently, only approaches that use simulation-optimization (meta-heuristic) techniques have considered stochastic task durations.
Whether finding optimal solutions, e.g., via stochastic linear programming, can practically be applied is questionable, as incorporating stochastic properties into the problem formulation will likely increase the problem complexity.
As, e.g., the benchmarking results from \cite{havur_benchmarking_2022} revealed that solving a deterministic scheduling problem for a process instance with 64 activities and 32 resources could already not finish within 2 hours of time, one can expect that for a stochastic formulation only small problem types will be solvable in a feasible amount of time.
%
%
%
Recent advancements in deep reinforcement learning have shown remarkable results for different kinds of optimization problems \citep{li_deep_2018}.
Future work could explore how deep reinforcement learning approaches can be applied to solve optimization problems in the context of business process management.

\noindent\textbf{Research Direction:} Stochastic approaches better represent real-world scenarios but can often not be solved sufficiently. Deep reinforcement learning could be employed to change this \citep{li_deep_2018}. Process models model uncertainty explicitly by using XOR-Gateways. How this modeling can be used to further improve stochastic approaches should be researched. A process focus can help improve the reinforcement learning approaches, especially for scheduling problems.

%
\section{Related work}
\label{sec:related_work}
%
%
%
Multiple surveys have focused on the topic of business process optimization \citep{kubrak_prescriptive_2022, pufahl_automatic_2021} taking a specific method from BPM and analyzing how it has been used to improve business processes. This SLR offers a broader scope in terms of optimization techniques since it is not limited to approaches prevalent in the BPM world. By incorporating OR into the research questions, the optimization-driven research area of OR as a whole is analyzed for possible optimization approaches that could advance automated business process optimization in the future. In the following, we discuss further research directions that are related but beyond the scope of this SLR.

\subsection{Prescriptive Process Monitoring}
Optimizing business processes based on previously obtained process traces has also been addressed through prescriptive process monitoring. Prescriptive process monitoring is defined as \enquote{a family of techniques to optimize the performance of a business process by triggering interventions at runtime.} \citep{shoush_prescriptive_2022}.
\citet{kubrak_prescriptive_2022} identified that predictive process mining methods can be categorized into two optimization objectives, i.e., optimizing the probability of a positive process outcome (categorical outcome), and optimizing the process efficiency (temporal outcome), e.g., cycling time.
They further found that prescriptions could either relate to the control flow perspective, e.g., which task to execute next, or to the resource perspective, e.g., which resource to choose from the pool of possible resources to execute a task. In contrast to our work, the review only focuses on optimization affecting process enactment but does not consider any optimization affecting the design phase at all. Further, it focuses on the BPM field and does not consider the combination with OR optimization approaches.
\subsection{Resource allocation}
Resource allocation strategies for business processes during the enactment phase have been surveyed in  \citet{pufahl_automatic_2021}. They distinguish between approaches that use a process model or process data as input for allocation decisions.
The majority of the results show that the most common resource allocation strategies are based on either simple, non-learnable decision rules or machine learning approaches such as genetic algorithms.
Non-learnable decision rule approaches concentrate mainly on selecting the best-fitting task-resource matching.
Only a few works solve allocation problems with the use of linear programming. The strict focus on resource allocation in the survey emphasizes the importance of resources for business process optimization. Yet, important factors for resource allocation, such as the used modeling of resources, have been excluded. As we have shown in this SLR, automated business process improvement can be achieved through multiple different optimization approaches, and thus, a narrow focus purely on resource allocation algorithms is not sufficient to cover the topic.  
\subsection{Scientific, grid and cloud computing}
Workflow technologies have also found applications to conduct large-scale scientific calculations.
In this field workflows are also known as \enquote{scientific workflows}.
Computing scientific workflows by means of distributed computers has been a subject of the grid- and cloud-computing disciplines. Grid computing is a term from the 90s from which cloud computing has emerged when new technologies have raised different questions \citep{foster_cloud_2008}.
Workflows in grid and cloud computing often also use more advanced workflow patterns than only directed acyclic graphs, such as choices or loops \citep{yu_taxonomy_2005}.
The disciplines are also concerned with optimizing the execution of the workflows, such as allocating workflow instances or tasks to optimal resources \citep{yu_taxonomy_2005}.
\cite{barker_scientific_2008} point out that while business and scientific workflows \enquote{began from the same ground}, \enquote{they each have their own domain-specific requirements, and therefore need separate consideration}.
%
While cloud computing typically only considers computing resources such as VMs \citep{liu_survey_2020}, BPM has a wider perception of resources including, i.a., process participants, software systems, and equipment \citep{dumas_fundamentals_2018} that might unveil behavior to be considered during optimization.
\section{Discussion and Conclusion}
We have analyzed publications that show how a combination of BPM and OR can enable the automatic optimization of business processes. By deriving our research questions and review protocol from the integrated model that was designed in advance (see \autoref{fig:bpm_or}), we are able to draw a holistic picture of the current state of research at the intersection of these two optimization focused research fields. 
\label{sec:discussion_conclusion}
\subsection{Discussion and Limitations}
%
%
%
%
%
The survey aims to illuminate how research has combined techniques from BPM and OR and identifies research gaps and potentials for future research.
Despite a careful study design, the results are exposed to several limitations:
\begin{enumerate}
    \item BPM and OR are both wide fields with 
    rather specific terms that were not explicitly part of the search strings.
    \citet{laengle_forty_2017}, for example, list the top 40 keywords used in the European Journal of Operations Research. However, as using all keywords was not feasible for this research, this study is limited to a subset that the authors deemed most relevant.
    \item The identified research gaps are derived from the analyzed publications, i.e., must have been stated in one of the publications. 
    \item The field of cloud computing has intensively applied workflow management technologies. Some optimization approaches in this field might be applicable to business process management. However, due to the size of the research field and the potentially bias to be introduced by that, the authors decided to exclude the field.
    Future research might investigate whether optimization techniques from cloud computing can be applied to business processes.
\end{enumerate}

\subsection{Conclusion}
This work analyzes existing research at the intersection of OR and BPM.
By resolving \textbf{RQ1--RQ3}, we examined the range of optimization purposes addressed in existing literature, how business process models and event logs have been conceptualized as mathematical models, and which solution techniques have been applied for finding optimal or near-optimal solutions.
We derived the findings by addressing each research question individually. Further, we identified open research gaps for future work.

%
%
\noindent\textit{\textbf{RQ1}: Which optimization purposes are addressed when techniques from OR are applied to business processes?}
Existing works address optimization problems in either the design or enactment phase, with the majority  addressing the enactment phase and, in particular, resource allocation and scheduling problems.
No approach has yet investigated a holistic approach concerned with both phases, mutual goals of both phases, even or contradictory objectives.
Furthermore, most works focus on a single optimization objective.
Future work should consider multi-objective optimization from all life-cycle phases. 

%
\noindent\textit{\textbf{RQ2}: How are business processes modeled or mined for the application of OR techniques?} 
Important information from process models or event logs is often disregarded when constructing an optimization model.
Results show that in the control flow perspective, most works do not consider complex workflow patterns, such as the exclusive choice pattern.
From the resource perspective, information on the effect of resources on the execution performance of tasks or processes is often missing.
While the event log is mostly used to structure control-flow data, no such consensus is found for the resource and data perspective. To combine several perspectives into one model a standardized data structure could be helpful. This can also enable the automatic generation of simulation models, which can help to optimize business processes from several perspectives.

%
\noindent\textit{\textbf{RQ3}: Which optimization techniques are used to optimize business processes?}
Optimization techniques strongly depend on their mathematical model.
As many approaches set up simplified mathematical models, optimal solution techniques have been widely used.
However, as simple models often neglect implications from the control flow, resource, and data perspective, solving larger, more holistic models with optimal algorithms is infeasible.
Future approaches should use solution techniques that can be applied to more complicated models that entail information from all process perspectives and work with the stochastic nature of problems. An automatic transformation from a process model and possible additional process perspectives into an e.g. optimization model for scheduling could help bridge this gap between process and optimization model.
Applying deep neural networks for prediction and decision-making might be a promising research field that has mostly been neglected in the surveyed publications.

This SLR identified important research gaps for the combination of OR and BPM. While the scope of OR is on optimizing specific mathematical problems, business process optimization thrives to enable the application of optimization approaches to general process models. This SLR shows that a better specification of the interface between OR and BPM is required. BPM-driven data models try to reflect the real world in as much detail as possible. Solution techniques from OR need abstracted data to find (optimal) solutions in a reasonable time. We see the design of dynamic systems that can leverage a continuous interplay of optimization algorithms and active BPM to achieve performance improvement as the core challenge when it comes to achieving automatic business process optimization. This study provides an overview of the current state of the art. The identified research gaps point to areas that must be addressed to achieve this goal.
%
%
%

\setstretch{2.0}

\printbibliography

@inproceedings{liu_survey_2020,
	title = {A Survey of Modern Scientific Workflow Scheduling Algorithms and Systems in the Era of Big Data},
	doi = {10.1109/SCC49832.2020.00026},
	eventtitle = {2020 {IEEE} International Conference on Services Computing ({SCC})},
	pages = {132--141},
	booktitle = {Services Computing},
	author = {Liu, Junwen and Lu, Shiyong and Che, Dunren},
	date = {2020-11},
}

@inproceedings{hirsch_optimization_2012,
	title = {On the optimization of information workflow},
	volume = {20},
	url = {https://www.scopus.com/inward/record.uri?eid=2-s2.0-84892739995&doi=10.1007%2f978-1-4614-3906-6_3&partnerID=40&md5=a56ed6251293cc9df5cad0b2d93d849d},
	doi = {10.1007/978-1-4614-3906-6_3},
	eventtitle = {International Conference on the Dynamics of Information Systems},
	pages = {43--65},
	booktitle = {Springer Proceedings in Mathematics and Statistics},
	author = {Hirsch, M.J. and Ortiz-Peña, H. and Nagi, R. and Sudit, M. and Stotz, A.},
	date = {2012},
}

@article{barba_automatic_2013,
	title = {Automatic generation of optimized business process models from constraint-based specifications},
	volume = {22},
	issn = {02188430 ({ISSN})},
	url = {https://www.scopus.com/inward/record.uri?eid=2-s2.0-84887865402&doi=10.1142%2fS0218843013500093&partnerID=40&md5=8eee74b2b16b212b761b84db049914e8},
	doi = {10.1142/S0218843013500093},
	number = {2},
	journaltitle = {International Journal of Cooperative Information Systems},
	shortjournal = {Int. J. Coop. Inf. Syst.},
	author = {Barba, I. and Del Valle, C. and Weber, B. and Jiménez, A.},
	date = {2013},
}

@article{barba_user_2013,
	title = {User recommendations for the optimized execution of business processes},
	volume = {86},
	issn = {0169023X ({ISSN})},
	url = {https://www.scopus.com/inward/record.uri?eid=2-s2.0-84878115641&doi=10.1016%2fj.datak.2013.01.004&partnerID=40&md5=ac9cc36b94e07e617668613f92855242},
	doi = {10.1016/j.datak.2013.01.004},
	pages = {61--84},
	journaltitle = {Data and Knowledge Engineering},
	shortjournal = {Data Knowl Eng},
	author = {Barba, I. and Weber, B. and Del Valle, C. and Jiménez-Ramírez, A.},
	date = {2013},
}

@inproceedings{jimenez-ramirez_generating_2013,
	title = {Generating Multi-Objective Optimized Business Process Enactment Plans},
	booktitle = {Advanced Information Systems Engineering},
	author = {Jiménez-Ramírez, A. and Barba, I. and Del Valle, C. and Weber, B.},
	date = {2013},
	doi = {10.1007/978-3-642-38709-8_7},
}

@inproceedings{camargo_learning_2019,
	location = {Cham},
	title = {Learning Accurate {LSTM} Models of Business Processes},
		doi = {10.1007/978-3-030-26619-6_19},
	pages = {286--302},
	booktitle = {Business Process Management},
		author = {Camargo, Manuel and Dumas, Marlon and González-Rojas, Oscar},
		date = {2019},
	langid = {english},
}

@article{van_der_aalst_business_2016,
	title = {Business Process Management},
	volume = {58},
	doi = {10.1007/s12599-015-0409-x},
	pages = {1--6},
	number = {1},
	journaltitle = {Business \& Information Systems Engineering},
	shortjournal = {Bus Inf Syst Eng},
	author = {van der Aalst, Wil M. P. and La Rosa, Marcello and Santoro, Flávia Maria},
	urldate = {2022-11-14},
	date = {2016},
	langid = {english},
}

@article{van_der_aalst_business_2013,
	title = {Business process management: a comprehensive survey},
	volume = {2013},
	issn = {2090-7680},
	url = {https://research.tue.nl/nl/publications/9661d40e-359b-43f4-b72c-dd4ee0ce6f1a},
	doi = {10.1155/2013/507984},
	shorttitle = {Business process management},
	journaltitle = {Software Engineering},
	author = {van der Aalst, Wil M. P.},
	urldate = {2022-11-14},
	date = {2013},
	langid = {english},
}

@article{van_der_aalst_workflow_2003,
	title = {Workflow Patterns},
	volume = {14},
	doi = {https://doi.org/10.1023/A:1022883727209},
	pages = {5--51},
	journaltitle = {Distributed and Parallel Databases},
	author = {van der Aalst, Wil M.P. and Ter Hofstede, Arthur H.M. and Kiepuszewski, B. and Barros, A. P.},
	date = {2003},
}

@inproceedings{poll_process_2018,
	location = {Cham},
	title = {Process Forecasting: Towards Proactive Business Process Management},
	isbn = {978-3-319-98648-7},
	doi = {10.1007/978-3-319-98648-7_29},
	pages = {496--512},
	booktitle = {Business Process Management},
	author = {Poll, Rouven and Polyvyanyy, Artem and Rosemann, Michael and Röglinger, Maximilian and Rupprecht, Lea},
	date = {2018},
	langid = {english},
}

@article{kinast_combing_2022,
	title = {Combing metaheuristics and process mining: Improving cobot placement in a combined cobot assignment and job shop scheduling problem},
	volume = {200},
	issn = {1877-0509},
	url = {https://www.sciencedirect.com/science/article/pii/S1877050922003933},
	doi = {10.1016/j.procs.2022.01.384},
	series = {3rd International Conference on Industry 4.0 and Smart Manufacturing},
	shorttitle = {Combing metaheuristics and process mining},
	pages = {1836--1845},
	journaltitle = {Procedia Computer Science},
	shortjournal = {Procedia Computer Science},
	author = {Kinast, Alexander and Doerner, Karl F. and Rinderle-Ma, Stefanie},
	urldate = {2022-10-20},
	date = {2022-01-01},
	langid = {english},
}

@article{havur_benchmarking_2022,
	title = {Benchmarking Answer Set Programming systems for resource allocation in business processes},
	volume = {205},
	issn = {09574174 ({ISSN})},
		doi = {10.1016/j.eswa.2022.117599},
	journaltitle = {Expert Systems with Applications},
	shortjournal = {Expert Sys Appl},
	author = {Havur, G. and Cabanillas, C. and Polleres, A.},
	date = {2022},
}

@inproceedings{havur_resource_2016,
	title = {Resource Allocation with Dependencies in Business Process Management Systems},
	volume = {260},
	isbn = {18651348 ({ISSN}); 9783319454672 ({ISBN})},
	url = {https://www.scopus.com/inward/record.uri?eid=2-s2.0-84988638277&doi=10.1007%2f978-3-319-45468-9_1&partnerID=40&md5=083dd828e34c0c36864867c1226d56ab},
	eventtitle = {{BPM} 2016},
	booktitle = {Business Process Management Forum},
	publisher = {Springer Verlag},
	author = {Havur, G. and Cabanillas, C. and Mendling, J. and Polleres, A.},
	editorb = {{Pastor O.} and {Loos P.} and {Rosa M.L.}},
	editorbtype = {redactor},
	date = {2016},
	doi = {10.1007/978-3-319-45468-9_1},
}

@inproceedings{hsieh_location-aware_2016,
	title = {Location-aware workflow scheduling in supply chains based on multi-agent systems},
	isbn = {9781467396066 ({ISBN})},
	url = {https://doi.org/10.1109/TAAI.2015.7407087},
	doi = {10.1109/TAAI.2015.7407087},
	eventtitle = {Conference on Technologies and Applications of Artificial Intelligence ({TAAI})},
	pages = {441--448},
	booktitle = {Technol. Appl. Artif. Intell.},
	publisher = {Institute of Electrical and Electronics Engineers Inc.},
	author = {Hsieh, F.-S.},
	date = {2016},
}

@article{xu_resource_2016,
	title = {Resource management for business process scheduling in the presence of availability constraints},
	volume = {7},
	issn = {2158656X ({ISSN})},
	url = {https://doi.org/10.1145/2990197},
	doi = {10.1145/2990197},
	number = {3},
	journaltitle = {{ACM} Transactions on Management Information Systems},
	shortjournal = {{ACM} Trans. Manage. Inf. Syst.},
	author = {Xu, J. and Liu, C. and Zhao, X. and Yongchareon, S. and Ding, Z.},
	date = {2016},
}

@inproceedings{ismaili-alaoui_business_2018,
	title = {Business process instances scheduling with human resources based on event priority determination},
	volume = {872},
	isbn = {18650929 ({ISSN})},
	url = {https://www.scopus.com/inward/record.uri?eid=2-s2.0-85063722204&doi=10.1007%2f978-3-319-96292-4_10&partnerID=40&md5=caa594f768ebea9e210f23dc1c076352},
	series = {Communications in Computer and Information Science},
	booktitle = {Big Data, Cloud and Applications},
	publisher = {Springer Verlag},
	author = {Ismaili-Alaoui, Abir and Benali, Khalid and Baïna, Karim and Baïna, Jamal},
	date = {2018},
	doi = {10.1007/978-3-319-96292-4_10},
}

@inproceedings{zheng_enterprise_2018,
	title = {Enterprise workflow modeling based on priced timed petri nets},
	isbn = {9781538658505 ({ISBN})},
	url = {https://www.scopus.com/inward/record.uri?eid=2-s2.0-85051490719&doi=10.1109%2fCSCloud%2fEdgeCom.2018.00027&partnerID=40&md5=ca99da68ff70d0bd071880ad9cab1357},
	doi = {10.1109/CSCloud/EdgeCom.2018.00027},
	eventtitle = {5th {IEEE} International Conference on Cyber Security and Cloud Computing ({CSCloud})/2018 4th {IEEE} International Conference on Edge Computing and Scalable Cloud ({EdgeCom})},
	pages = {106--110},
	booktitle = { Cyber Secur. Cloud Comput. {IEEE} Int. Conf. Edge Comput. Scalable Cloud},
	author = {Zheng, H. and Li, L. and Wang, B. and Ruan, T.},
	editor = {{Qiu M.}},
	date = {2018},
}

@article{subramaniam_improving_2007,
	title = {Improving process models by discovering decision points},
	volume = {32},
	issn = {03064379 ({ISSN})},
	url = {https://www.scopus.com/inward/record.uri?eid=2-s2.0-34347332203&doi=10.1016%2fj.is.2006.11.001&partnerID=40&md5=3da531caacb667a03dd8c3bd77b37e96},
	doi = {10.1016/j.is.2006.11.001},
	pages = {1037--1055},
	number = {7},
	journaltitle = {Information Systems},
	shortjournal = {Inf. Syst.},
	author = {Subramaniam, S. and Kalogeraki, V. and Gunopulos, D. and Casati, F. and Castellanos, M. and Dayal, U. and Sayal, M.},
	date = {2007},
}

@article{avanes_adaptive_2008,
	title = {Adaptive workflow scheduling under resource allocation constraints and network dynamics},
	volume = {1},
	issn = {21508097 ({ISSN})},
	url = {https://www.scopus.com/inward/record.uri?eid=2-s2.0-76349115519&doi=10.14778%2f1454159.1454238&partnerID=40&md5=60bb29eb42ecd647e70cf1c0f1cba8bd},
	doi = {10.14778/1454159.1454238},
	pages = {1631--1367},
	number = {2},
	journaltitle = {{VLDB} },
	shortjournal = {Proc. {VLDB} Endow.},
	author = {Avanes, A. and Freytag, J.-C.},
	date = {2008},
}

@inproceedings{delias_joint_2008,
	title = {A joint optimization algorithm for dispatching tasks in agent-based workflow management systems},
	isbn = {9789898111388 ({ISBN}); 9789898111371 ({ISBN})},
	url = {https://www.scopus.com/inward/record.uri?eid=2-s2.0-55349143180&partnerID=40&md5=59d05724989e0797b224304a1b3fa8c1},
	eventtitle = {Enterprise Information Systems},
	pages = {199--206},
	booktitle = {Enterprise Information Systems},
	author = {Delias, P. and Doulamis, A. and Matsatsinis, N.},
	date = {2008},
}

@inproceedings{antunes_solution_2019,
 author       = {Bianca B. P. Antunes and
                  Adrian Manresa and
                  Leonardo S. L. Bastos and
                  Janaina Figueira Marchesi and
                  Silvio Hamacher},
 title        = {A Solution Framework Based on Process Mining, Optimization, and Discrete-Event
                  Simulation to Improve Queue Performance in an Emergency Department},
  booktitle    = {Business Process Management Workshops },
  pages        = {583--594},
  year         = {2019},
   doi          = {10.1007/978-3-030-37453-2\_47}
}

@inproceedings{julia_p-time_2004,
	title = {A p-time hybrid Petri net model for the scheduling problem of Workflow Management Systems},
	volume = {5},
	doi = {10.1109/ICSMC.2004.1401315},
	booktitle = {Systems, Man and Cybernetics},
	pages = {4947--4952},
	author = {Julia, S. and De Oliveira, F.F.},
	date = {2004},
}

@article{choueiri_discovery_2021,
	title = {Discovery of path-attribute dependency in manufacturing environments: A process mining approach},
	volume = {61},
	issn = {02786125},
	url = {https://www.scopus.com/inward/record.uri?eid=2-s2.0-85113383945&doi=10.1016%2fj.jmsy.2021.08.005&partnerID=40&md5=fa1706da1857811ffd1d74fa4ec7b58d},
	doi = {10.1016/j.jmsy.2021.08.005},
	pages = {54--65},
	journaltitle = {Journal of Manufacturing Systems},
	author = {Choueiri, A.C. and Portela Santos, E.A.},
	date = {2021},
	note = {Publisher: Elsevier B.V.},
}

@article{zhao_entropy-based_2016,
	title = {An entropy-based clustering ensemble method to support resource allocation in business process management},
	volume = {48},
	url = {https://www.scopus.com/inward/record.uri?eid=2-s2.0-84940976747&doi=10.1007%2fs10115-015-0879-7&partnerID=40&md5=9e9604c610958974478f380b72640694},
	doi = {10.1007/s10115-015-0879-7},
	pages = {305--330},
	number = {2},
	journaltitle = {Knowledge and Information Systems},
	author = {Zhao, W. and Liu, H. and Dai, W. and Ma, J.},
	date = {2016},
}

@inproceedings{lee_clustering_2017,
	title = {Clustering and operation analysis for assembly blocks using process mining in shipbuilding industry},
	pages = {67--80},
    date = {2013},
	booktitle = {Asia Pacific Business Process Mgmt.},
	author = {Lee, D. and Park, J. and Pulshashi, I.R. and Bae, H.},
}

@article{arias_human_2018,
	title = {Human resource allocation or recommendation based on multi-factor criteria in on-demand and batch scenarios},
	volume = {12},
	doi = {10.1504/EJIE.2018.092009},
	pages = {364--404},
	number = {3},
	journaltitle = {European Journal of Industrial Engineering},
	author = {Arias, M. and Munoz-Gama, J. and Sepúlveda, M. and Miranda, J.C.},
	date = {2018},
}

@article{djedovic_innovative_2018,
	title = {Innovative Approach in Modeling Business Processes with a Focus on Improving the Allocation of Human Resources},
	volume = {2018},
	url = {https://www.scopus.com/inward/record.uri?eid=2-s2.0-85054416998&doi=10.1155%2f2018%2f9838560&partnerID=40&md5=8a451420c62769bb915ccd45912b5578},
	doi = {10.1155/2018/9838560},
	journaltitle = {Mathematical Problems in Engineering},
	author = {Djedovic, A. and Karabegovic, A. and Avdagic, Z. and Omanovic, S.},
	date = {2018},
}

@article{lee_dynamic_2019,
	title = {Dynamic human resource selection for business process exceptions},
	volume = {26},
	url = {https://www.scopus.com/inward/record.uri?eid=2-s2.0-85057988572&doi=10.1002%2fkpm.1591&partnerID=40&md5=19362fee043122722f75324102a79365},
	doi = {10.1002/kpm.1591},
	pages = {23--31},
	number = {1},
	journaltitle = {Knowl. and Process Management},
	author = {Lee, J. and Lee, S. and Kim, J. and Choi, I.},
	date = {2019},
}

@inproceedings{havur_history-aware_2019,
  author       = {Giray Havur and
                  Cristina Cabanillas},
  title        = {History-Aware Dynamic Process Fragmentation for Risk-Aware Resource
                  Allocation},
  booktitle    = {On the Move to Meaningful Internet Systems},
  pages        = {533--551},
   year         = {2019},
   doi          = {10.1007/978-3-030-33246-4\_33}
}

@inproceedings{park_prediction-based_2019,
	title = {Prediction-based resource allocation using {LSTM} and minimum cost and maximum flow algorithm},
	url = {https://www.scopus.com/inward/record.uri?eid=2-s2.0-85071182772&doi=10.1109%2fICPM.2019.00027&partnerID=40&md5=4c06d02dbb1afc20e8075022c5158021},
	doi = {10.1109/ICPM.2019.00027},
	pages = {121--128},
	booktitle = {Process Mining},
	author = {Park, G. and Song, M.},
	date = {2019},
}

@article{prokofyeva_clinical_2020,
	title = {Clinical pathways analysis of patients in medical institutions based on hard and fuzzy clustering methods},
	volume = {14},
	url = {https://www.scopus.com/inward/record.uri?eid=2-s2.0-85097677541&doi=10.17323%2f2587-814X.2020.1.19.31&partnerID=40&md5=2ca497b8401b7a6c150e6289cdaf6945},
	doi = {10.17323/2587-814X.2020.1.19.31},
	pages = {19--31},
	number = {1},
	journaltitle = {Business Informatics},
	author = {Prokofyeva, E.S. and Zaytsev, R.D.},
	date = {2020},
}

@article{li_mining_2021,
	title = {Mining emergency event logs to support resource allocation},
	url = {https://www.scopus.com/inward/record.uri?eid=2-s2.0-85116514490&doi=10.1587%2ftransinf.2021EDP7029&partnerID=40&md5=f618c101670a1c09c4131b1d845725c4},
	doi = {10.1587/transinf.2021EDP7029},
	pages = {1651--1660},
	number = {10},
	journaltitle = {Transactions on Information and Systems},
	author = {Li, H. and Liu, C. and Zeng, Q. and He, H. and Ren, C. and Wang, L. and Cheng, F.},
	date = {2021},
}

@article{yeon_experimental_2022,
	title = {Experimental Verification on Human-Centric Network-Based Resource Allocation Approaches for Process-Aware Information Systems},
	volume = {10},
	url = {https://www.scopus.com/inward/record.uri?eid=2-s2.0-85125307898&doi=10.1109%2fACCESS.2022.3152778&partnerID=40&md5=ace1fcff0c71fdd6ee11936c77b6f8a3},
	doi = {10.1109/ACCESS.2022.3152778},
	pages = {23342--23354},
	journaltitle = {{IEEE} Access},
	author = {Yeon, M.-S. and Lee, Y.-K. and Pham, D.-L. and Kim, K.P.},
	date = {2022},
}

@inproceedings{lopez-pintado_business_2022,
  author       = {Orlenys L{\'{o}}pez{-}Pintado and
                  Marlon Dumas},
  title        = {Business Process Simulation with Differentiated Resources: Does it
                  Make a Difference?},
  booktitle    = {Business Process Management},
  pages        = {361--378},
  year         = {2022},
   doi          = {10.1007/978-3-031-16103-2\_24},
 }

@article{ruschel_mining_2017,
	title = {Mining Shop-Floor Data for Preventive Maintenance Management: Integrating Probabilistic and Predictive Models},
	volume = {11},
	url = {https://www.scopus.com/inward/record.uri?eid=2-s2.0-85029846774&doi=10.1016%2fj.promfg.2017.07.234&partnerID=40&md5=7d8064fed017b2bd6500f266dae19d1b},
	doi = {10.1016/j.promfg.2017.07.234},
	pages = {1127--1134},
	journaltitle = {Procedia Manufacturing},
	author = {Ruschel, E. and Santos, E.A.P. and Loures, E.D.F.R.},
	date = {2017},
}

@article{senderovich_conformance_2016,
	title = {Conformance checking and performance improvement in scheduled processes: A queueing-network perspective},
	volume = {62},
	url = {https://www.scopus.com/inward/record.uri?eid=2-s2.0-84957900578&doi=10.1016%2fj.is.2016.01.002&partnerID=40&md5=66902b911ebb5df83b0867fb227271a0},
	doi = {10.1016/j.is.2016.01.002},
	pages = {185--206},
	journaltitle = {Information Systems},
	author = {Senderovich, A. and Weidlich, M. and Yedidsion, L. and Gal, A. and Mandelbaum, A. and Kadish, S. and Bunnell, C.A.},
	date = {2016},
	note = {Publisher: Elsevier Ltd},
}

@book{taha_operations_2017,
	location = {Harlow, {UNITED} {KINGDOM}},
	title = {Operations Research: an Introduction, Global Edition},
	isbn = {978-1-292-16556-1},
	url = {http://ebookcentral.proquest.com/lib/munchentech/detail.action?docID=5832625},
	shorttitle = {Operations Research},
	publisher = {Pearson Education, Limited},
	author = {Taha, Hamdy},
	urldate = {2022-12-01},
	date = {2017},
}

@inproceedings{zhao_optimization_2015,
	title = {The optimization of resource allocation based on process mining},
	volume = {9227},
	url = {https://www.scopus.com/inward/record.uri?eid=2-s2.0-84983593798&doi=10.1007%2f978-3-319-22053-6_38&partnerID=40&md5=72d7363c5a72cfbed4f9abf6bed1de8d},
	doi = {10.1007/978-3-319-22053-6_38},
	pages = {341--353},
	author = {Zhao, W. and Yang, L. and Liu, H. and Wu, R.},
  booktitle  = {Advanced Intelligent Computing Theories and Applications},
 date = 2015,
}

@inproceedings{van_der_aalst_business_2003,
	location = {Berlin, Heidelberg},
	title = {Business Process Management: A Survey},
	isbn = {978-3-540-44895-2},
	doi = {10.1007/3-540-44895-0_1},
	series = {Lecture Notes in Computer Science},
	shorttitle = {Business Process Management},
	pages = {1--12},
	booktitle = {Business Process Management},
	author = {van der Aalst, Wil M. P. and ter Hofstede, Arthur H. M. and Weske, Mathias},
	date = {2003},
	langid = {english},
}

@article{huang_reinforcement_2011,
	title = {Reinforcement learning based resource allocation in business process management},
	volume = {70},
	issn = {0169-023X},
	url = {https://www.sciencedirect.com/science/article/pii/S0169023X1000114X},
	doi = {10.1016/j.datak.2010.09.002},
	pages = {127--145},
	number = {1},
	journaltitle = {Data \& Knowledge Engineering},
	shortjournal = {Data \& Knowledge Engineering},
	author = {Huang, Zhengxing and van der Aalst, W. M. P. and Lu, Xudong and Duan, Huilong},
	urldate = {2022-12-23},
	date = {2011-01-01},
	langid = {english},
}

@techreport{kitchenham_guidelines_2007,
	title = {Guidelines for performing Systematic Literature Reviews in Software Engineering},
	volume = {2},
	author = {Kitchenham, Barbara and Charters, Stuart},
	date = {2007-01-01},
  institution = {University of Durham} 
}

@article{cho_evidence-based_2019,
	title = {An Evidence-Based Decision Support Framework for Clinician Medical Scheduling},
	volume = {7},
		doi = {10.1109/ACCESS.2019.2894116},
	pages = {15239--15249},
	journaltitle = {{IEEE} Access},
	author = {Cho, Minsu and Song, Minseok and Yoo, Sooyoung and Reijers, Hajo A.},
	date = {2019},
}

@article{schulte_elastic_2015,
	title = {Elastic Business Process Management: State of the art and open challenges for {BPM} in the cloud},
	volume = {46},
	issn = {0167-739X},
	url = {https://www.sciencedirect.com/science/article/pii/S0167739X1400168X},
	doi = {10.1016/j.future.2014.09.005},
	shorttitle = {Elastic Business Process Management},
	pages = {36--50},
	journaltitle = {Future Generation Computer Systems},
	shortjournal = {Future Generation Computer Systems},
	author = {Schulte, Stefan and Janiesch, Christian and Venugopal, Srikumar and Weber, Ingo and Hoenisch, Philipp},
	urldate = {2023-01-21},
	date = {2015-05-01},
	langid = {english},
}

@book{hillier_introduction_2020,
	title = {Introduction to Operations Research},
	isbn = {978-1-260-57587-3},
	pagetotal = {964},
	publisher = {{McGraw}-Hill Education},
	author = {Hillier, Frederick S. and Lieberman, Gerald J.},
	date = {2020},
	langid = {english},
}

@inproceedings{peters_resource_2021,
	title = {Resource Optimization in Business Processes},
	doi = {10.1109/EDOC52215.2021.00021},
	eventtitle = {2021 {IEEE} 25th International Enterprise Distributed Object Computing Conference ({EDOC})},
	pages = {104--113},
	booktitle = {Enterprise Distributed Object Computing Conference},
	author = {Peters, S. P. F. and Dijkman, R. M. and Grefen, P. W. P. J.},
	date = {2021-10},
}

@article{pufahl_automatic_2021,
	title = {Automatic Resource Allocation in Business Processes: A Systematic Literature Survey},
	url = {http://arxiv.org/abs/2107.07264},
	shorttitle = {Automatic Resource Allocation in Business Processes},
	author = {Pufahl, Luise and Ihde, Sven and Stiehle, Fabian and Weske, Mathias and Weber, Ingo},
	urldate = {2023-01-30},
 journal      = {CoRR},
  volume       = {abs/2107.07264},
	date = {2021},
	langid = {english},
	eprinttype = {arxiv},
	eprint = {2107.07264 [cs]},
}

@article{julia_real_2008,
	title = {Real time scheduling of Workflow Management Systems based on a p-time Petri net model with hybrid resources},
	volume = {16},
	issn = {1569-190X},
	url = {https://www.sciencedirect.com/science/article/pii/S1569190X08000166},
	doi = {10.1016/j.simpat.2008.01.006},
	pages = {462--482},
	number = {4},
	journaltitle = {Simulation Modelling Practice and Theory},
	shortjournal = {Simulation Modelling Practice and Theory},
	author = {Julia, Stéphane and de Oliveira, Fernanda Francielle and Valette, Robert},
	urldate = {2023-02-13},
	date = {2008-04-01},
	langid = {english},
}

@inproceedings{medeiros_constraint_2017,
	location = {Porto, Portugal},
	title = {Constraint Analysis based on Energetic Reasoning Applied to the Problem of Real Time Scheduling of Workflow Management Systems},
		doi = {10.5220/0006275903730380},
	shorttitle = {Constraint Analysis based on Energetic Reasoning Applied to the Problem of Real Time Scheduling of Workflow Management Systems},
	eventtitle = {19th International Conference on Enterprise Information Systems},
	pages = {373--380},
    date = {2017},
	booktitle = { Enterprise Information Systems},
		author = {Medeiros, Flávio Félix and Julia, Stéphane},
}

@inproceedings{satyal_ab-bpm_2017,
	location = {Cham},
	title = {{AB}-{BPM}: Performance-Driven Instance Routing for Business Process Improvement},
	isbn = {978-3-319-65000-5},
	doi = {10.1007/978-3-319-65000-5_7},
	series = {Lecture Notes in Computer Science},
	shorttitle = {{AB}-{BPM}},
	pages = {113--129},
	booktitle = {Business Process Management},
	author = {Satyal, Suhrid and Weber, Ingo and Paik, Hye-young and Di Ciccio, Claudio and Mendling, Jan},
	date = {2017},
	langid = {english},
}

@inproceedings{satyal_ab_2018,
	location = {Cham},
	title = {{AB} Testing for Process Versions with Contextual Multi-armed Bandit Algorithms},
	isbn = {978-3-319-91563-0},
	doi = {10.1007/978-3-319-91563-0_2},
	series = {Lecture Notes in Computer Science},
	eventtitle = {International Conference on Advanced Information Systems Engineering},
	pages = {19--34},
	booktitle = {Advanced Information Systems Engineering},
	author = {Satyal, Suhrid and Weber, Ingo and Paik, Hye-young and Di Ciccio, Claudio and Mendling, Jan},
	date = {2018},
	langid = {english},
}

@book{eiselt_operations_2010,
	location = {Berlin/Heidelberg, {GERMANY}},
	title = {Operations Research: A Model-Based Approach},
	isbn = {978-3-642-10326-1},
	url = {http://ebookcentral.proquest.com/lib/munchentech/detail.action?docID=3065283},
	shorttitle = {Operations Research},
	publisher = {Springer Berlin / Heidelberg},
	author = {Eiselt, H. A. and Sandblom, Carl-Louis},
	urldate = {2023-02-08},
	date = {2010},
}

@article{ruschel_establishment_2020,
	title = {Establishment of maintenance inspection intervals: an application of process mining techniques in manufacturing},
	volume = {31},
	issn = {1572-8145},
	url = {https://doi.org/10.1007/s10845-018-1434-7},
	doi = {10.1007/s10845-018-1434-7},
	shorttitle = {Establishment of maintenance inspection intervals},
	pages = {53--72},
	number = {1},
	journaltitle = {Journal of Intelligent Manufacturing},
	shortjournal = {J Intell Manuf},
	author = {Ruschel, Edson and Santos, Eduardo Alves Portela and Loures, Eduardo de Freitas Rocha},
	urldate = {2023-02-15},
	date = {2020-01-01},
	langid = {english},
}

@inproceedings{kurscheidt_netto_enabling_2021,
	location = {Cham},
	title = {Enabling the Use of Shop Floor Information for Multi-criteria Decision Making in Maintenance Prediction},
	isbn = {978-3-030-78570-3},
	doi = {10.1007/978-3-030-78570-3_33},
	pages = {435--447},
	booktitle = {Industrial Engineering and Operations Management},
 author = {Kurscheidt Netto, Rolando J. and de F. R. Loures, Eduardo and dos Santos, Eduardo A. P.},
	date = {2021},
	langid = {english},
}

@inproceedings{djedovic_optimization_2016,
	title = {Optimization of business processes by automatic reallocation of resources using the genetic algorithm},
	doi = {10.1109/BIHTEL.2016.7775724},
	eventtitle = {2016 {XI} International Symposium on Telecommunications ({BIH}℡)},
	pages = {1--7},
	booktitle = {Telecommunications},
	author = {Djedović, Almir and Žunić, Emir and Avdagić, Zikrija and Karabegović, Almir},
	date = {2016-10},
}

@article{reveliotis_real-time_2016,
	title = {Real-Time Management of Complex Resource Allocation Systems: Necessity, Achievements and Further Challenges},
	volume = {41},
	pages = {147--158},
	journaltitle = {Annual Reviews in Control},
	author = {Reveliotis, Spyros},
	date = {2016},
	langid = {english},
}

@inproceedings{pla_multi-attribute_2012,
	title = {Multi-Attribute Auction Mechanism for Supporting Resource Allocation in Business Process Enactment},
	url = {https://ebooks.iospress.nl/doi/10.3233/978-1-61499-096-3-228},
	doi = {10.3233/978-1-61499-096-3-228},
	pages = {228--239},
	booktitle = {Starting {AI} Researchers’ Symposium},
	author = {Pla, Albert and Lopez, Beatriz and Murillo, Javier},
	urldate = {2023-03-08},
	date = {2012},
	note = {Publisher: {IOS} Press},
}

@inproceedings{bae_--fly_2015,
	location = {Cham},
	title = {On-the-Fly Performance-Aware Human Resource Allocation in the Business Process Management Systems Environment Using Naïve Bayes},
	volume = {219},
	isbn = {978-3-319-19508-7 978-3-319-19509-4},
	url = {https://link.springer.com/10.1007/978-3-319-19509-4_6},
	eventtitle = {Asia-Pacific Conference on Business Process Management},
	pages = {70--80},
	booktitle = {Asia Pacific Business Process Management},
	author = {Wibisono, Arif and Nisafani, Amna Shifia and Bae, Hyerim and Park, You-Jin},
	urldate = {2023-03-15},
	date = {2015},
	langid = {english},
	doi = {10.1007/978-3-319-19509-4_6},
}

@article{bolsinger_process_2015,
	title = {Process improvement through economically driven routing of instances},
	volume = {21},
	issn = {1463-7154},
	url = {https://doi.org/10.1108/BPMJ-02-2014-0011},
	doi = {10.1108/BPMJ-02-2014-0011},
	pages = {353--378},
	number = {2},
	journaltitle = {Business Process Management Journal},
	author = {Bolsinger, Manuel and Elsäßer, Anna and Helm, Caroline and Röglinger, Maximilian},
	urldate = {2023-03-14},
	date = {2015-01-01},
}

@article{xu_incorporating_2013,
	title = {Incorporating structural improvement into resource allocation for business process execution planning},
	volume = {25},
	issn = {1532-0634},
	url = {https://onlinelibrary.wiley.com/doi/abs/10.1002/cpe.2855},
	doi = {10.1002/cpe.2855},
	pages = {427--442},
	number = {3},
	journaltitle = {Concurrency and Computation: Practice and Experience},
	author = {Xu, Jiajie and Liu, Chengfei and Zhao, Xiaohui and Ding, Zhiming},
	urldate = {2023-03-22},
	date = {2013},
	langid = {english},
}

@article{bai_risk_2013,
	title = {On Risk Management with Information Flows in Business Processes},
	volume = {24},
	issn = {1047-7047},
	url = {https://pubsonline.informs.org/doi/abs/10.1287/isre.1120.0450},
	doi = {10.1287/isre.1120.0450},
	pages = {731--749},
	number = {3},
	journaltitle = {Inf. Syst. Research},
	author = {Bai, Xue and Krishnan, Ramayya and Padman, Rema and Wang, Harry Jiannan},
	urldate = {2023-03-23},
	date = {2013-09},
}

@inproceedings{senderovich_learning_2019,
	title = {Learning Scheduling Models from Event Data},
	volume = {29},
	url = {https://ojs.aaai.org/index.php/ICAPS/article/view/3504},
	doi = {10.1609/icaps.v29i1.3504},
	pages = {401--409},
	booktitle = {Proceedings of the International Conference on Automated Planning and Scheduling},
	author = {Senderovich, Arik and Booth, Kyle E. C. and Beck, J. Christopher},
	urldate = {2023-03-23},
	date = {2019},
	langid = {english},
}

@article{wang_trust-based_2013,
	title = {Trust-based workflow refactoring for concurrent scheduling in service-oriented environment},
	volume = {25},
	issn = {1532-0634},
	url = {https://onlinelibrary.wiley.com/doi/abs/10.1002/cpe.2989},
	doi = {10.1002/cpe.2989},
	pages = {1879--1893},
	number = {13},
	journaltitle = {Concurrency and Computation: Practice and Experience},
	author = {Wang, Mingzhong and Zhang, Xuyun and Zhu, Liehuang and Liao, Lejian},
	urldate = {2023-03-23},
	date = {2013},
	langid = {english},
}

@inproceedings{yang_approach_2012,
	location = {Berlin, Heidelberg},
	title = {An Approach to Recommend Resources for Business Processes},
	isbn = {978-3-642-33618-8},
	doi = {10.1007/978-3-642-33618-8_88},
	series = {Lecture Notes in Computer Science},
	eventtitle = {{OTM} Confederated International Conferences "On the Move to Meaningful Internet Systems"},
	pages = {662--665},
	booktitle = {On the Move to Meaningful Internet Systems},
	author = {Yang, Hedong and Wen, Lijie and Liu, Yingbo and Wang, Jianmin},
	date = {2012},
	langid = {english},
}

@article{liu_semi-automatic_2008,
	title = {A semi-automatic approach for workflow staff assignment},
	volume = {59},
	issn = {01663615},
	url = {https://linkinghub.elsevier.com/retrieve/pii/S0166361507001832},
	doi = {10.1016/j.compind.2007.12.002},
	pages = {463--476},
	number = {5},
	journaltitle = {Computers in Industry},
	shortjournal = {Computers in Industry},
	author = {Liu, Yingbo and Wang, Jianmin and Yang, Yun and Sun, Jiaguang},
	urldate = {2023-04-03},
	date = {2008-05},
	langid = {english},
}

@inproceedings{combi_task_2006,
	title = {Task Scheduling for a Temporal Workflow Management System},
	doi = {10.1109/TIME.2006.26},
	eventtitle = {International Workshop on Temporal Representation and Reasoning},
	pages = {61--68},
	booktitle = { Temporal Representation and Reasoning},
	author = {Combi, C. and Pozzi, G.},
	date = {2006-06},
}

@article{lin_practical_2004,
	title = {A practical scheduling method based on workflow management technology},
	volume = {24},
	issn = {1433-3015},
	url = {https://doi.org/10.1007/s00170-003-1799-3},
	doi = {10.1007/s00170-003-1799-3},
	pages = {919--924},
	number = {11},
	journaltitle = {The International Journal of Advanced Manufacturing Technology},
	shortjournal = {Int J Adv Manuf Technol},
	author = {Lin, Huiping and Fan, Yushun and Loiacono, Eleanor T.},
	urldate = {2023-04-12},
	date = {2004-12-01},
	langid = {english},
}

@inproceedings{niedermann_design-time_2010,
	title = {Design-Time Process Optimization through Optimization Patterns and Process Model Matching},
	doi = {10.1109/CEC.2010.9},
	eventtitle = {2010 {IEEE} 12th Conference on Commerce and Enterprise Computing},
	pages = {48--55},
	booktitle = {Commerce and Enterprise Computing},
	author = {Niedermann, Florian and Radeschutz, Sylvia and Mitschang, Bernhard},
	date = {2010-11},
}

@inproceedings{shoush_prescriptive_2022,
	location = {Cham},
	title = {Prescriptive Process Monitoring Under Resource Constraints: A Causal Inference Approach},
	isbn = {978-3-030-98581-3},
	doi = {10.1007/978-3-030-98581-3_14},
	series = {Lecture Notes in Business Information Processing},
	shorttitle = {Prescriptive Process Monitoring Under Resource Constraints},
	pages = {180--193},
	booktitle = {Process Mining Workshops},
	author = {Shoush, Mahmoud and Dumas, Marlon},
	date = {2022},
	langid = {english},
}

@article{kubrak_prescriptive_2022,
	title = {Prescriptive process monitoring: Quo vadis?},
	volume = {8},
	doi = {10.7717/peerj-cs.1097},
	shorttitle = {Prescriptive process monitoring},
	pages = {e1097},
	journaltitle = {{PeerJ} Computer Science},
	shortjournal = {{PeerJ} Comput. Sci.},
	author = {Kubrak, Kateryna and Milani, Fredrik and Nolte, Alexander and Dumas, Marlon},
	urldate = {2023-04-27},
	date = {2022-09-29},
	langid = {english}
}

@book{dumas_fundamentals_2018,
	location = {Berlin, Heidelberg, {GERMANY}},
	title = {Fundamentals of Business Process Management},
	isbn = {978-3-662-56509-4},
	url = {http://ebookcentral.proquest.com/lib/munchentech/detail.action?docID=6315916},
	publisher = {Springer Berlin / Heidelberg},
	author = {Dumas, Marlon and La Rosa, Marcello and Mendling, Jan and Reijers, Hajo A.},
	urldate = {2023-05-03},
	date = {2018},
}

@inproceedings{pufahl_evaluating_2015,
	location = {Cham},
	title = {Evaluating the Performance of a Batch Activity in Process Models},
	isbn = {978-3-319-15895-2},
	doi = {10.1007/978-3-319-15895-2_24},
	series = {Lecture Notes in Business Information Processing},
	eventtitle = {International Conference on Business Process Management},
	pages = {277--290},
	booktitle = {Business Process Management Workshops},
	author = {Pufahl, Luise and Bazhenova, Ekaterina and Weske, Mathias},
	editor = {Fournier, Fabiana and {Jan Mendling}},
	date = {2015},
	langid = {english},
}

@article{pflug_application_2016,
	title = {Application of Dynamic Instance Queuing to Activity Sequences in Cooperative Business Process Scenarios},
	volume = {25},
	issn = {0218-8430},
	url = {https://www.worldscientific.com/doi/10.1142/S0218843016500027},
	doi = {10.1142/S0218843016500027},
	pages = {1650002},
	number = {1},
	journaltitle = { Cooperative Information Systems},
	shortjournal = {Int. J. Coop. Info. Syst.},
	author = {Pflug, Johannes and Rinderle-Ma, Stefanie},
	urldate = {2023-05-05},
	date = {2016-03},
}

@article{zeng_effective_2005,
	title = {Effective Role Resolution in Workflow Management},
	volume = {17},
	issn = {1091-9856},
	url = {https://pubsonline.informs.org/doi/abs/10.1287/ijoc.1040.0067},
	doi = {10.1287/ijoc.1040.0067},
	pages = {374--387},
	number = {3},
	journaltitle = {{INFORMS} Journal on Computing},
	author = {Zeng, Daniel D. and Zhao, J. Leon},
	urldate = {2023-05-17},
	date = {2005-08},
}

@article{li_deep_2018,
	title = {Deep Reinforcement Learning: An Overview},
	url = {http://arxiv.org/abs/1701.07274},
	shorttitle = {Deep Reinforcement Learning},
	number = {{arXiv}:1701.07274},
	publisher = {{arXiv}},
	author = {Li, Yuxi},
	urldate = {2023-07-12},
	date = {2018-11-25},
	langid = {english},
	eprinttype = {arxiv},
	eprint = {1701.07274},
}

@incollection{dijkman_1st_2023,
	location = {Cham},
	title = {1st International Workshop on Data-Driven Business Process Optimization},
	volume = {460},
	isbn = {978-3-031-25383-6},
	series = {Lecture Notes in Business Information Processing},
	shorttitle = {Business Process Management Workshops},
	pages = {336},
	booktitle = {Business Process Management Workshops},
	publisher = {Springer},
	author = {Dijkman, Remco and Senderovich, Arik and van Jaarsveld, Willem},
	date = {2023},
	langid = {english},
}

@article{hsieh_hybrid_2017,
	title = {A hybrid and scalable multi-agent approach for patient scheduling based on Petri net models},
	volume = {47},
	issn = {1573-7497},
	url = {https://doi.org/10.1007/s10489-017-0935-y},
	doi = {10.1007/s10489-017-0935-y},
	pages = {1068--1086},
	number = {4},
	journaltitle = {Applied Intelligence},
	shortjournal = {Appl Intell},
	author = {Hsieh, Fu-Shiung},
	urldate = {2023-08-18},
	date = {2017-12-01},
	langid = {english},
}

@article{comuzzi_ant-colony_2019,
	title = {Ant-Colony Optimisation for Path Recommendation in Business Process Execution},
	volume = {8},
	issn = {1861-2040},
	url = {https://doi.org/10.1007/s13740-018-0099-x},
	doi = {10.1007/s13740-018-0099-x},
	pages = {113--128},
	number = {2},
	journaltitle = {Journal on Data Semantics},
	shortjournal = {J Data Semant},
	author = {Comuzzi, Marco},
	urldate = {2023-08-17},
	date = {2019-06-01},
	langid = {english},
}

@inproceedings{weinzierl_prescriptive_2020,
	location = {Cham},
	title = {Prescriptive Business Process Monitoring for Recommending Next Best Actions},
	isbn = {978-3-030-58638-6},
	doi = {10.1007/978-3-030-58638-6_12},
	series = {Lecture Notes in Business Information Processing},
	pages = {193--209},
	booktitle = {Business Process Management Forum},
	author = {Weinzierl, Sven and Dunzer, Sebastian and Zilker, Sandra and Matzner, Martin},
	date = {2020},
	langid = {english},
}

@article{van_der_aalst_re-engineering_2001,
	title = {Re-engineering knock-out processes},
	volume = {30},
	issn = {0167-9236},
	url = {https://www.sciencedirect.com/science/article/pii/S0167923600001366},
	doi = {10.1016/S0167-9236(00)00136-6},
	pages = {451--468},
	number = {4},
	journaltitle = {Decision Support Systems},
	shortjournal = {Decision Support Systems},
	author = {van der Aalst, Wil M. P.},
	urldate = {2023-08-17},
	date = {2001-03-01},
}

@inproceedings{yaghoubi_tuning_2018,
	location = {New York, {NY}, {USA}},
	title = {Tuning Concurrency of the Business Process by Dynamic Programming},
	isbn = {978-1-4503-5414-1},
	url = {https://dl.acm.org/doi/10.1145/3185089.3185090},
	doi = {10.1145/3185089.3185090},
	series = {{ICSCA} '18},
	eventtitle = {International Conference on Software and Computer Applications},
	pages = {1--5},
	booktitle = {Software and Computer Applications},
	publisher = {Association for Computing Machinery},
	author = {Yaghoubi, Mehdi and Zahedi, Morteza},
	urldate = {2023-09-11},
	date = {2018-02-08},
}

@article{laengle_forty_2017,
	title = {Forty years of the European Journal of Operational Research: A bibliometric overview},
	volume = {262},
	issn = {0377-2217},
	url = {https://www.sciencedirect.com/science/article/pii/S0377221717303600},
	doi = {10.1016/j.ejor.2017.04.027},
	shorttitle = {Forty years of the European Journal of Operational Research},
	pages = {803--816},
	number = {3},
	journaltitle = {European Journal of Operational Research},
	shortjournal = {European Journal of Operational Research},
	author = {Laengle, Sigifredo and Merigó, José M. and Miranda, Jaime and Słowiński, Roman and Bomze, Immanuel and Borgonovo, Emanuele and Dyson, Robert G. and Oliveira, José Fernando and Teunter, Ruud},
	urldate = {2023-10-10},
	date = {2017-11-01},
}

@inproceedings{bando_automatic_2022,
	title = {Automatic Generation of Optimization Model using Process Mining and Petri Nets for Optimal Motion Planning of 6-{DOF} Manipulators},
	url = {https://ieeexplore.ieee.org/document/9982201},
	doi = {10.1109/IROS47612.2022.9982201},
	pages = {11767--11772},
	booktitle = {Intelligent Robots and Syst. },
	author = {Bando, Takuma and Nishi, Tatsushi and Alam, Md Moktadir and Liu, Ziang and Fujiwara, Tomofumi},
	urldate = {2023-10-16},
	date = {2022-10},
}

@inproceedings{van_der_aalst_change_2015,
	title = {Change your history: Learning from event logs to improve processes},
	url = {https://ieeexplore.ieee.org/document/7230925},
	doi = {10.1109/CSCWD.2015.7230925},
	shorttitle = {Change your history},
	eventtitle = {2015 {IEEE} 19th International Conference on Computer Supported Cooperative Work in Design ({CSCWD})},
	pages = {7--12},
	booktitle = {Computer Supported Cooperative Work in Design},
	author = {van der Aalst, Wil M. P. and Low, Wei Zhe and Wynn, Moe T. and ter Hofstede, Arthur H.M.},
	urldate = {2023-10-18},
	date = {2015-05},
}

@article{rismanchian_process_2017,
	title = {Process Mining–Based Method of Designing and Optimizing the Layouts of Emergency Departments in Hospitals},
	volume = {10},
	issn = {1937-5867},
	url = {https://doi.org/10.1177/1937586716674471},
	doi = {10.1177/1937586716674471},
	pages = {105--120},
	number = {4},
	journaltitle = { Health Environments Research \& Design Journal},
	shortjournal = {{HERD}},
	author = {Rismanchian, Farhood and Lee, Young Hoon},
	urldate = {2023-10-18},
	date = {2017-07-01},
	langid = {english},
}

@inproceedings{guastalla_workshift_2023,
	location = {Singapore, Singapore},
	title = {Workshift Scheduling Using Optimization and Process Mining Techniques: An Application in Healthcare},
	series = {{WSC} '22},
	shorttitle = {Workshift Scheduling Using Optimization and Process Mining Techniques},
	pages = {1116--1127},
	booktitle = {Proceedings of the Winter Simulation Conference},
	publisher = {{IEEE} Press},
	author = {Guastalla, Alberto and Sulis, Emilio and Aringhieri, Roberto and Branchi, Stefano and Di Francescomarino, Chiara and Ghidini, Chiara},
	urldate = {2023-10-25},
	date = {2023-03-02},
}

@inproceedings{neubauer_resource_2022,
	location = {Cham},
	title = {Resource Allocation Optimization in Business Processes Supported by Reinforcement Learning and Process Mining},
	isbn = {978-3-031-21686-2},
	doi = {10.1007/978-3-031-21686-2_40},
	pages = {580--595},
	booktitle = {Intelligent Systems},
	author = {Neubauer, Thais Rodrigues and da Silva, Valdinei Freire and Fantinato, Marcelo and Peres, Sarajane Marques},
	date = {2022},
	langid = {english},
}

@article{kumbhar_digital_2023,
	title = {A digital twin based framework for detection, diagnosis, and improvement of throughput bottlenecks},
	volume = {66},
	issn = {0278-6125},
	url = {https://www.sciencedirect.com/science/article/pii/S0278612522002151},
	doi = {10.1016/j.jmsy.2022.11.016},
	pages = {92--106},
	journaltitle = {Journal of Manufacturing Systems},
	shortjournal = {Journal of Manufacturing Systems},
	author = {Kumbhar, Mahesh and Ng, Amos H. C. and Bandaru, Sunith},
	urldate = {2023-10-27},
	date = {2023-02-01},
}

@article{liu_workflow_2012,
	title = {Workflow simulation for operational decision support using event graph through process mining},
	volume = {52},
	issn = {0167-9236},
	url = {https://www.sciencedirect.com/science/article/pii/S0167923611002077},
	doi = {10.1016/j.dss.2011.11.003},
	pages = {685--697},
	number = {3},
	journaltitle = {Decision Support Systems},
	shortjournal = {Decision Support Systems},
	author = {Liu, Ying and Zhang, Hui and Li, Chunping and Jiao, Roger Jianxin},
	urldate = {2023-10-27},
	date = {2012-02-01},
}

@article{bae_planning_2014,
	title = {Planning of business process execution in Business Process Management environments},
	volume = {268},
		doi = {10.1016/j.ins.2013.12.061},
	series = {New Sensing and Processing Technologies for Hand-based Biometrics Authentication},
	pages = {357--369},
	journaltitle = {Inf. Sciences},
	shortjournal = {Information Sciences},
	author = {Bae, Hyerim and Lee, Sanghyup and Moon, Ilkyeong},
	urldate = {2023-11-09},
	date = {2014-06-01},
}

@article{delias_optimizing_2011,
	title = {Optimizing Resource Conflicts in Workflow Management Systems},
	volume = {23},
	issn = {1558-2191},
	url = {https://ieeexplore.ieee.org/document/5518766},
	doi = {10.1109/TKDE.2010.113},
	pages = {417--432},
	number = {3},
	journaltitle = {{IEEE} Transactions on Knowledge and Data Engineering},
	author = {Delias, Pavlos and Doulamis, Anastasios and Doulamis, Nikolaos and Matsatsinis, Nikolaos},
	urldate = {2023-11-09},
	date = {2011-03},
}

@article{yu_taxonomy_2005,
	title = {A Taxonomy of Workflow Management Systems for Grid Computing},
	volume = {3},
	issn = {1572-9184},
	url = {https://doi.org/10.1007/s10723-005-9010-8},
	doi = {10.1007/s10723-005-9010-8},
	pages = {171--200},
	number = {3},
	journaltitle = {Journal of Grid Computing},
	shortjournal = {J Grid Computing},
	author = {Yu, Jia and Buyya, Rajkumar},
	urldate = {2023-11-11},
	date = {2005-09-01},
	langid = {english},
}

@inproceedings{foster_cloud_2008,
	location = {Austin, {TX}, {USA}},
	title = {Cloud Computing and Grid Computing 360-Degree Compared},
	isbn = {978-1-4244-2860-1},
	url = {http://ieeexplore.ieee.org/document/4738445/},
	doi = {10.1109/GCE.2008.4738445},
	eventtitle = {2008 Grid Computing Environments Workshop},
	pages = {1--10},
	booktitle = {Grid Comp. Environments Workshop},
	publisher = {{IEEE}},
	author = {Foster, Ian and Zhao, Yong and Raicu, Ioan and Lu, Shiyong},
	urldate = {2023-11-11},
	date = {2008-11},
	langid = {english},
}

@inproceedings{barker_scientific_2008,
	location = {Berlin, Heidelberg},
	title = {Scientific Workflow: A Survey and Research Directions},
		doi = {10.1007/978-3-540-68111-3_78},
	shorttitle = {Scientific Workflow},
	pages = {746--753},
	booktitle = {Parallel Processing and Applied Mathematics},
		author = {Barker, Adam and van Hemert, Jano},
		date = {2008},
	langid = {english},
}

@article{tucek_main_2015,
	title = {The Main Reasons for Implementing {BPM} in Czech Companies},
	volume = {7},
	issn = {1804171X, 18041728},
	url = {http://www.cjournal.cz/index.php?hid=clanek&cid=200},
	doi = {10.7441/joc.2015.03.09},
	pages = {126--142},
	number = {3},
	journaltitle = {Journal of Competitiveness},
	shortjournal = {{JOC}},
	author = {Tucek, David},
	urldate = {2023-11-13},
	date = {2015-09-30},
	langid = {english},
}

@article{halawa_integrated_2021,
	title = {Integrated framework of process mining and simulation–optimization for pod structured clinical layout design},
	volume = {185},
	issn = {0957-4174},
	url = {https://www.sciencedirect.com/science/article/pii/S0957417421010800},
	doi = {10.1016/j.eswa.2021.115696},
	pages = {115696},
	journaltitle = {Expert Systems with Applications},
	shortjournal = {Expert Systems with Applications},
	author = {Halawa, Farouq and Chalil Madathil, Sreenath and Khasawneh, Mohammad T.},
	urldate = {2023-11-15},
	date = {2021-12-15},
}

@inproceedings{cho_new_2017,
	location = {Cham},
	title = {A New Framework for Defining Realistic {SLAs}: An Evidence-Based Approach},
		doi = {10.1007/978-3-319-65015-9_2},
		shorttitle = {A New Framework for Defining Realistic {SLAs}},
	pages = {19--35},
	booktitle = {Business Process Management Forum},
		author = {Cho, Minsu and Song, Minseok and Müller, Carlos and Fernandez, Pablo and del-Río-Ortega, Adela and Resinas, Manuel and Ruiz-Cortés, Antonio},
		date = {2017},
	langid = {english},
}

@article{low_revising_2016,
	title = {Revising history for cost-informed process improvement},
	volume = {98},
	issn = {1436-5057},
	url = {https://doi.org/10.1007/s00607-015-0478-1},
	doi = {10.1007/s00607-015-0478-1},
	pages = {895--921},
	number = {9},
	journaltitle = {Computing},
	shortjournal = {Computing},
	author = {Low, W. Z. and van den Broucke, S. K. L. M. and Wynn, M. T. and ter Hofstede, A. H. M. and De Weerdt, J. and van der Aalst, W. M. P.},
	urldate = {2023-11-15},
	date = {2016-09-01},
	langid = {english},
}

@inproceedings{liu_q-learning_2015,
	location = {Berlin, Heidelberg},
	title = {Q-learning Algorithm for Task Allocation Based on Social Relation},
	isbn = {978-3-662-46170-9},
	doi = {10.1007/978-3-662-46170-9_5},
	series = {Communications in Computer and Information Science},
	pages = {49--58},
	booktitle = {Process-Aware Systems},
	publisher = {Springer},
	author = {Liu, Xingmei and Chen, Jian and Ji, Yu and Yu, Yang},
	date = {2015},
	langid = {english},
}

@inproceedings{xu_performance_2012,
	location = {Xiangtan, China},
	title = {A Performance Analysis on Task Allocation Using Social Context},
	eventtitle = {Conference on Cloud and Green Computing},
	booktitle = {Cloud and Green Computing},
	author = {Xu, Jiaxing and Huang, Zhenguang and Yu, Yang and Pan, Maolin},
pages        = {637--644},
	date = {2012},
	langid = {english},
}

@article{ng_business_2018,
	title = {Business process optimization using the ant colony system},
	volume = {39},
	issn = {1099-1468},
	url = {https://onlinelibrary.wiley.com/doi/abs/10.1002/mde.2933},
	doi = {10.1002/mde.2933},
	pages = {629--637},
	number = {6},
	journaltitle = {Managerial and Decision Economics},
	author = {Ng, C.y.},
	urldate = {2023-11-15},
	date = {2018},
	langid = {english},
}

@inproceedings{alonso_workflow_2007,
	location = {Berlin, Heidelberg},
	title = {Workflow Management Systems + Swarm Intelligence = Dynamic Task Assignment for Emergency Management Applications},
	url = {http://link.springer.com/10.1007/978-3-540-75183-0_10},
	pages = {125--140},
	booktitle = {Business Process Management},
	author = {Reijers, Hajo A. and Jansen-Vullers, Monique H. and Zur Muehlen, Michael and Appl, Winfried},
	urldate = {2023-11-29},
	date = {2007},
	langid = {english},
	doi = {10.1007/978-3-540-75183-0_10},
}

@article{banaszak_workflows_2003,
	title = {Workflows Management for Project-Driven Manufacturing},
	volume = {36},
	issn = {14746670},
	url = {https://linkinghub.elsevier.com/retrieve/pii/S1474667017377492},
	doi = {10.1016/S1474-6670(17)37749-2},
	pages = {149--154},
	number = {3},
	journaltitle = {{IFAC} Proceedings Volumes},
	shortjournal = {{IFAC} Proceedings Volumes},
	author = {Banaszak, Zbigniew A.},
	urldate = {2023-11-29},
	date = {2003-04},
	langid = {english},
}

@inproceedings{di_cunzolo_combining_2023,
	location = {Cham},
	title = {Combining Process Mining and Optimization: A Scheduling Application in Healthcare},
	doi = {10.1007/978-3-031-25383-6_15},
	pages = {197--209},
	booktitle = {Business Process Management Workshops},
	author = {Di Cunzolo, Matteo and Guastalla, Alberto and Aringhieri, Roberto and Sulis, Emilio and Amantea, Ilaria Angela and Ronzani, Massimiliano and Di Francescomarino, Chiara and Ghidini, Chiara and Fonio, Paolo and Grosso, Marco},
		date = {2023},
	langid = {english},
}

@article{delias_formulating_2023,
	title = {Formulating the potentials of clustering of event data over multiple entities for decision support: a network embeddings approach},
	volume = {0},
	issn = {1246-0125},
	url = {https://doi.org/10.1080/12460125.2023.2263684},
	doi = {10.1080/12460125.2023.2263684},
	shorttitle = {Formulating the potentials of clustering of event data over multiple entities for decision support},
	pages = {1--23},
	number = {0},
	journaltitle = {Journal of Decision Systems},
	author = {Delias, Pavlos and Grigori, Daniela},
	urldate = {2023-11-30},
	date = {2023},
}

@article{song_identifying_2022,
	title = {Identifying a Minimum Sequence of High-Level Changes Between Workflows},
	volume = {15},
	issn = {1939-1374, 2372-0204},
	url = {https://ieeexplore.ieee.org/document/9335480/},
	doi = {10.1109/TSC.2021.3054036},
	pages = {2425--2438},
	number = {4},
	journaltitle = {{IEEE} Transactions on Services Computing},
	shortjournal = {{IEEE} Trans. Serv. Comput.},
	author = {Song, Wei and Chen, Fangfei and Jacobsen, Hans-Arno and Zhang, Chengzhen},
	urldate = {2023-12-04},
	date = {2022-07-01},
	langid = {english},
}

@article{liu_multi-level_2022,
	title = {Multi-level Team Assignment in Social Business Processes: An Algorithm and Simulation Study},
	volume = {24},
	issn = {1572-9419},
	url = {https://doi.org/10.1007/s10796-021-10211-y},
	doi = {10.1007/s10796-021-10211-y},
	shorttitle = {Multi-level Team Assignment in Social Business Processes},
	pages = {1949--1969},
	number = {6},
	journaltitle = {Information Systems Frontiers},
	shortjournal = {Inf Syst Front},
	author = {Liu, Rong and Kumar, Akhil and Lee, Juhnyoung},
	urldate = {2023-12-04},
	date = {2022-12-01},
	langid = {english},
}

@article{juan_review_2015,
	title = {A review of simheuristics: Extending metaheuristics to deal with stochastic combinatorial optimization problems},
	volume = {2},
	issn = {2214-7160},
	url = {https://www.sciencedirect.com/science/article/pii/S221471601500007X},
	doi = {10.1016/j.orp.2015.03.001},
	shorttitle = {A review of simheuristics},
	pages = {62--72},
	journaltitle = {Operations Research Perspectives},
	shortjournal = {Operations Research Perspectives},
	author = {Juan, Angel A. and Faulin, Javier and Grasman, Scott E. and Rabe, Markus and Figueira, Gonçalo},
	urldate = {2023-12-13},
	date = {2015-12-01},
}

@article{zhu_tale_2020,
	title = {A tale of two databases: the use of Web of Science and Scopus in academic papers},
	volume = {123},
	issn = {1588-2861},
	url = {https://doi.org/10.1007/s11192-020-03387-8},
	doi = {10.1007/s11192-020-03387-8},
	shorttitle = {A tale of two databases},
	pages = {321--335},
	number = {1},
	journaltitle = {Scientometrics},
	shortjournal = {Scientometrics},
	author = {Zhu, Junwen and Liu, Weishu},
	urldate = {2024-01-17},
	date = {2020-04-01},
	langid = {english},
}

@article{van_der_aalst_process_2012,
	title = {Process Mining: Overview and Opportunities},
	volume = {3},
	issn = {2158-656X, 2158-6578},
	url = {https://dl.acm.org/doi/10.1145/2229156.2229157},
	doi = {10.1145/2229156.2229157},
	shorttitle = {Process Mining},
	pages = {1--17},
	number = {2},
	journaltitle = {{ACM} Transactions on Management Information Systems},
	shortjournal = {{ACM} Trans. Manage. Inf. Syst.},
	author = {van der Aalst, Wil M. P.},
	urldate = {2024-01-26},
	date = {2012-07},
	langid = {english},
}

@incollection{vom_brocke_business_2010,
	location = {Berlin, Heidelberg},
	title = {Business Process Analytics},
	isbn = {978-3-642-01981-4 978-3-642-01982-1},
	url = {http://link.springer.com/10.1007/978-3-642-01982-1_7},
	pages = {137--157},
	booktitle = {Handbook on Business Process Management 2},
	author = {Zur Muehlen, Michael and Shapiro, Robert},
	urldate = {2024-01-26},
	date = {2010},
	langid = {english},
	doi = {10.1007/978-3-642-01982-1_7},
}

@inproceedings{van_der_aalst_business_2015,
	author       = {Wil M. P. van der Aalst},
  editor       = {Jan vom Brocke and
                  Michael Rosemann},
  title        = {Business Process Simulation Survival Guide},
  booktitle    = {Handbook on Business Process Management 1, Introduction, Methods,
                  and Information Systems, 2nd Ed},
  series       = {International Handbooks on Information Systems},
  pages        = {337--370},
  publisher    = {Springer},
  year         = {2015},
  url          = {https://doi.org/10.1007/978-3-642-45100-3\_15},
  doi          = {10.1007/978-3-642-45100-3\_15},
  bibsource    = {dblp computer science bibliography, https://dblp.org}
}

@article{amaran_simulation_2016,
	title = {Simulation optimization: a review of algorithms and applications},
	volume = {240},
	issn = {1572-9338},
	url = {https://doi.org/10.1007/s10479-015-2019-x},
	doi = {10.1007/s10479-015-2019-x},
	shorttitle = {Simulation optimization},
	pages = {351--380},
	number = {1},
	journaltitle = {Annals of Operations Research},
	shortjournal = {Ann Oper Res},
	author = {Amaran, Satyajith and Sahinidis, Nikolaos V. and Sharda, Bikram and Bury, Scott J.},
	urldate = {2024-04-05},
	date = {2016-05-01},
	langid = {english},
}

@article{likert_technique_1932,
	title = {A technique for the measurement of attitudes},
	volume = {22  140},
	pages = {55--55},
	journaltitle = {Archives of Psychology},
	author = {Likert, R.},
	date = {1932},
}
\end{document}